%% file: paper.tex
\documentclass[a4paper,useAMS,usenatbib,usegraphicx]{mn2e}
\input{psfig.sty}
\input{macros}
\begin{document}
\setlength{\hbadness}{10000}
%%%%%%%%%%%%%%%%%%%%%%%%%%%%%%%%%%%%%%%%%%%%%%%%%%%%%%%%%%%%%%%%%%%%%%%%%%

\title[Satellite Kinematics III]
      {Satellite Kinematics III: Halo Masses of Central Galaxies in SDSS}
\author[More et al.]
       {Surhud More$^{1,2}$ \thanks{E-mail: surhud@kicp.uchicago.edu}
        \thanks{KICP fellow}, 
        Frank C. van den Bosch$^{1,3}$, 
        Marcello Cacciato$^{1,4}$ \thanks{Minerva fellow}, \and 
        Ramin Skibba$^{1,5}$, 
        H. J. Mo$^{6}$, 
        Xiaohu Yang$^{7}$ \\
   $^{1}$Max Planck Institute for Astronomy, Koenigstuhl,17, D69117,
          Heidelberg, Germany.\\
   $^{2}$Kavli Institute for Cosmological Physics, University of Chicago,
	  933 East 56th Street, Chicago, IL 60637, USA \\
   $^{3}$Department of Physics and Astronomy, University of Utah, 115
	  South 1400 East, Salt Lake City, UT 84112-0830 \\
   $^{4}$Racah Institute of Physics, The Hebrew University, Jerusalem
	  91904, Israel \\
   $^{5}$Steward Observatory, University of Arizona, 933 N. Cherry
	  Avenue, Tucson, AZ 85721, USA \\
   $^{6}$Department of Astronomy, University of Massachussets, Amherst, MA
	  010039305, USA\\
   $^{7}$Shanghai Astronomical Observatory, Nandan Road 80, Shanghai
	  200030, China}

%%%%%%%%%%%%%%%%%%%%%%%%%%%%%%%%%%%%%%%%%%%%%%%%%%%%%%%%%%%%%%%%%%%%%%%%%%

\date{}

\maketitle

\label{firstpage}

%%%%%%%%%%%%%%%%%%%%%%%%%%%%%%%%%%%%%%%%%%%%%%%%%%%%%%%%%%%%%%%%%%%%%%%%%%

\begin{abstract}
  We use the kinematics of satellite galaxies that orbit around the
  central galaxy in a dark matter halo to infer the scaling relations
  between halo mass and central galaxy properties. Using galaxies from
  the Sloan Digital Sky Survey, we investigate the halo
  mass$-$luminosity relation (MLR) and the halo mass$-$stellar mass
  relation (MSR) of central galaxies. In particular, we focus on the
  dependence of these scaling relations on the colour of the central
  galaxy. We find that red central galaxies on average occupy more
  massive haloes than blue central galaxies of the same
  luminosity. However, at fixed stellar mass there is no appreciable
  difference in the average halo mass of red and blue centrals,
  especially for $M_*\lta 10^{10.5}\Msunhh$. This indicates that
  stellar mass is a better indicator of halo mass than
  luminosity. Nevertheless, we find that the scatter in halo masses at
  fixed stellar mass is non-negligible for both red and blue centrals.
  It increases as a function of stellar mass for red centrals but
  shows a fairly constant behaviour for blue centrals. We compare the
  scaling relations obtained in this paper with results from other
  independent studies of satellite kinematics, with results from a
  SDSS galaxy group catalog, from galaxy-galaxy weak lensing
  measurements, and from subhalo abundance matching studies. Overall,
  these different techniques yield MLRs and MSRs in fairly good
  agreement with each other (typically within a factor of two),
  indicating that we are converging on an accurate and reliable
  description of the galaxy-dark matter connection. We briefly discuss
  some of the remaining discrepancies among the various methods.
\end{abstract}

%%%%%%%%%%%%%%%%%%%%%%%%%%%%%%%%%%%%%%%%%%%%%%%%%%%%%%%%%%%%%%%%%%%%%%%%%%


%%%%%%%%%%%%%%%%%%%%%%%%%%%%%%%%%%%%%%%%%%%%%%%%%%%%%%%%%%%%%%%%%%%%%%%%%%
\section{Introduction} 
\label{sec:intro} 

The growth of structure in the Universe is predominantly driven by
dark matter.  The fluctuations in the dark matter density field grow
under the action of gravity and form a web-like structure. Galaxies
form as baryon condensates at the density peaks of this cosmic web.
Understanding the connection between the distribution of galaxies and
the underlying distribution of dark matter is crucial to understand
the physics of galaxy formation. This galaxy-dark matter connection is
often expressed in terms of the scaling relations between the
properties of galaxies and the mass of the dark matter halo in which
they reside. Reliable measurements of the dark matter halo mass are
essential to quantify these scaling relations. This can be
accomplished with the help of numerous methods. These include
techniques that are primarily used on individual systems such as
rotation curves \citep[e.g.,][]{Rubin1982}, strong lensing of
background galaxies \citep[e.g.,][]{Gavazzi2007}, and X-ray emision
from hot gas in clusters \citep[e.g.,][]{Rykoff2008,Dai2007}. With the
advent of large scale galaxy redshift surveys, substantial progress
has been made with methods that allow the inference of the dark matter
halo masses in a statistical sense, e.g.  the average halo mass as a
function of various properties of galaxies.  Such methods include the
modelling of the clustering of galaxies \citep[e.g., ][]{Yang2003,
Zehavi2004, Zehavi2005, Tinker2005, Collister2005, Skibba2006,
Bosch2007, Brown2008, Skibba2009},
galaxy-galaxy weak lensing \citep[e.g.,][]{Seljak2000, McKay2001,
Mandelbaum2006, Parker2007, Mandelbaum2008, Schulz2009} and a
combination of the two \citep[e.g.,][]{Yoo2006, Cacciato2009, Li2009}.

The satellite galaxies that orbit within the dark matter haloes of
their central galaxies are also excellent probes of the dark matter
halo mass. Their kinematics reflect the depth of the dark matter
potential well they orbit. The number of satellites in massive systems
like clusters is large enough to obtain a reliable measure of their
kinematics and hence the halo mass \citep[e.g.,][]{Carlberg1996,
Carlberg1997}.  However, in low mass systems, where only a handful of
satellites can be detected per central, one has to adopt a stacking
procedure to quantify the kinematics of satellites
\citep{Erickson1987, Zaritsky1993, Zaritsky1994, Zaritsky1997}.
Central galaxies with similar properties (e.g. luminosity) are stacked
together and the velocity information of their satellites is combined
to obtain a quantitative measure of the kinematics of the satellites.
Various studies of the kinematics of satellite galaxies have now been
carried out using large redshift surveys to improve the sample size of
satellites \citep{McKay2002, Brainerd2003, Prada2003, Conroy2005,
Becker2007, Conroy2007, Norberg2008}. The sample sizes in these
studies have been limited by their use of strict isolation criteria to
identify centrals and satellites, specifically designed to avoid
misidentifications.  \citet{vdb04} devised relaxed selection criteria
which were iteratively adapted to the luminosity of central galaxies
to circumvent this problem. Their criteria improved the sample size by
nearly an order of magnitude over studies that use strict isolation
criteria while still maintaining low levels of contamination. The
improved statistics have warranted a better study of the systematics
and selection effects that bias the kinematic measurements
\citep{Norberg2008}.

In More, van den Bosch \& Cacciato (2009b; hereafter Paper I), we
showed that if the relation between the halo mass and the stacking
property has a non-negligible scatter then the kinematics of the
satellites of the stacked system can be difficult to interpret. This
issue has been neglected by most previous studies. We presented a new
method to infer both the average halo mass and the scatter in halo
masses as a function of the property used to stack the central
galaxies. In More et al. (2009a; hereafter Paper II), this method was
applied to galaxies from the Sloan Digital Sky Survey (York et al.
2000; hereafter SDSS) to infer the halo mass$-$luminosity relation
of central galaxies (hereafter MLR).  It was found that both the
average and the scatter of the MLR of central galaxies increases with
the luminosity of the central galaxy.

The scatter in the MLR is an interesting quantity, as it is related to
the stochasticity in galaxy formation. Two important, related
questions are: (i) what is the physical origin of this stochasticity,
and (ii) what galaxy property is most closely related to the mass of
the halo in which it resides (i.e., shows the least amount of scatter
at a given halo mass). The answer to (i) yields valuable insight into
the physics of galaxy formation, while the answer to (ii) identifies
the optimal galaxy property to trace the cosmic density field. It is
well known that galaxies of the same stellar mass may have very
different luminosities, even after correction for dust
extinction. Galaxies with younger stellar populations will typically
be bluer and more luminous than galaxies of the same stellar mass, but
with an older stellar population. It may well be that the stellar mass
of a central galaxy is a better indicator of halo mass than its
luminosity, in which case the halo mass$-$stellar mass relation of
central galaxies (hereafter MSR) will have less scatter than the MLR,
and the scatter in the MLR will be correlated with the color of the
central galaxy. Obviously, since we lack a complete theory of galaxy
formation, it may also be that the opposite holds, and that the
scatter in the MLR is actually less than that in the MSR. In this
paper, we investigate these issue by measuring the kinematics of
satellite galaxies as functions of both the luminosity and stellar
mass of centrals split by colour into red and blue sub-samples.  Using
the methodology outlined in Paper~I, we use these to probe both the
means and scatters of the MLR and MSR of red and blue galaxies.

This paper is organized as follows. In Section~\ref{sec:data}, we
describe the data used in this paper. In Section~\ref{sec:methods}, we
explain our method of analysis. In particular, we describe the
procedure used to identify the centrals and the satellites, the
measurement of the kinematics of satellites, and the subsequent
modelling to determine the halo masses of central galaxies.  In
Section~\ref{sec:results} we present our results and compare them with
other independent studies. Finally, we summarize our findings in
Section~\ref{sec:summary}.

Throughout this paper we adopt the cosmological parameters supported
by the 3 year data release of WMAP \citep{Spergel2007}; $\Omega_{\rm
m}=0.238$, $\Omega_\Lambda=0.762$, $h=H_0/100 \kms \Mpc^{-1} =0.734$,
the spectral index of initial density fluctuations $n_{\rm s}=0.951$
and the normalization of the power spectrum of density fluctuations
$\sigma_8=0.744$. We use the symbol $M$ to refer to the mass of a dark
matter halo, which is defined as the mass enclosed within a spherical
overdensity $\delta\rho/\bar{\rho}=200$, where $\bar{\rho}$ denotes
the mean matter density of the universe.

\section{Data}
\label{sec:data}

We use data from the SDSS which is a joint five-passband ($u, g, r, i$
and $z$) imaging and medium resolution ($R \sim 1800$) spectroscopic
survey \citep{York2000}. More specifically, we use the New York
University Value Added Galaxy Catalogue \citep{Blanton2005b}, which is
based upon SDSS Data Release 4 \citep{Adelman-McCarthy2006} but
includes a set of significant improvements over the original
pipelines.  The magnitudes and colours of the galaxies are based upon
the standard SDSS Petrosian technique and have been k-corrected and
evolution corrected to $z=0.1$ using the method described in
\citet{Blanton2003a,Blanton2003b}. The notations $^{0.1}(g-r)$ and
$^{0.1}M_r - 5 \log h$ are used to denote the resulting $(g-r)$ colour
and the absolute magnitude of the galaxies.  From this catalogue, we
select all galaxies in the main galaxy sample with apparent magnitudes
less than $17.77$ that lie in an area where the redshift completeness
limit of the survey ${\cal C} > 0.8$. Next we construct a volume
limited sample that is complete in luminosity above a $^{0.1}r$-band
luminosity of $L_{\rm min}= 10^{9.5} \Lsunhh$\footnote{The
$^{0.1}r$-band magnitude of the Sun in the AB system equal to $4.76$
\citep{Blanton2003a} is used to convert the absolute magnitude of a
galaxy to its luminosity in units of $\Lsunhh$.}. The redshift range
that we adopt for this volume limited sample is $0.02 \le z \le
0.072$, which results in a total sample of $58,396$ galaxies.

Stellar masses are indicated by $M_*$ and are computed using the
relation between the stellar mass-to-light ratio and the $^{0.0}(g-r)$
colour provided by \citet{Bell2003}:
\begin{eqnarray}
\label{eq:bell2003}
\log\left[ \frac{M_*}{\Msunhh} \right] =& -0.306 + 1.097\,[^{0.0}(g-r)] -0.10
\nonumber \\                            & -0.4\,(^{0.0}M_r - 5 \log h-4.64).
\end{eqnarray}
Here $^{0.0}(g-r)$ and $^{0.0}M_r - 5 \log h$ denote the $(g-r)$
colour and the $r$-band absolute magnitude of galaxies k-corrected and
evolution corrected to $z=0.0$, $4.64$ is the $r$-band magnitude of
the Sun in the AB system, and the $-0.10$ term is a result of adopting
the \citet{Kroupa2001} initial mass function \citep[see][]{Borch2006}.
The typical uncertainties in the stellar masses obtained in this manner
are of the order of $\sim 0.1$ dex \citep{Bell2003}.

We classify galaxies to be red or blue based upon their bimodal
distribution in the $^{0.0}(g-r)$ colour-stellar mass plane. We use
the following separation criterion to demarcate the boundary between
red and blue galaxies in the colour-stellar mass plane (see
Appendix~\ref{app:a}):
\begin{equation}
^{0.0}(g-r)_{\rm cut} = 0.65 + 0.10\,\left(\log \left[ M_*/(h^{-2}M_\odot)
\right] - 10 .0 \right)\,.
\end{equation}

\section{Methodology}
\label{sec:methods}

\subsection{Selection criteria}
\label{sec:selncrit}

The first step towards measuring the kinematics of satellite galaxies
in the SDSS is to accurately identify central galaxies and their
associated satellites. For our analysis of the MLR we proceed as
follows. A galaxy is identified to be a central if it is brighter than
every other galaxy within a cylindrical volume specified by $R < \Rh$
and $|\dv|<\dvh$ centred on itself.  Here $R$ is the physical distance
from the galaxy under consideration projected on the sky and $\dv$ is
the line-of-sight (hereafter los) velocity difference between two
galaxies.  All galaxies that lie within a cylindrical volume specified
by $R < \Rs$ and $|\dv|<\dvs$ around a central galaxy, and that are
fainter than the central galaxy, are labelled to be its satellites.
The criteria used to select the sample of central and satellite
galaxies for the analysis of the MSR are almost identical, except that
in this case the central galaxy must have the largest stellar mass in
its cylindrical volume specified by $\Rh$ and $\dvh$.

The parameters $\Rh, \dvh, \Rs$ and $\dvs$ define the sizes of the
cylinders used to identify central galaxies and their
satellites. Contrary to most previous studies of satellite kinematics
\citep{McKay2002, Brainerd2003, Prada2003, Conroy2005, Conroy2007,
  Norberg2008}, we do not use fixed values for these parameters.
Rather, since halo mass is expected to be positively correlated with
the luminosity or stellar mass of the central galaxies, we scale the
selection parameters according to the property of the galaxy under
consideration. Following \citet{vdb04}, we adopt $\Rh = 0.8
\sigma_{200} h^{-1}\Mpc$, $\dvh = 1000 \sigma_{200} \kms$, $\Rs = 0.15
\sigma_{200} h^{-1}\Mpc$ and $\dvs = 4000 \kms$. Here $\sigma_{200}$
is the satellite velocity dispersion in units of $200\kms$, which we
parameterize as
\begin{equation}
 \sigma_{200} (\log \pten) = a + b~(\log \pten) + c~(\log \pten)^2.
\label{eq:sig200}
\end{equation}
where $\pten$ is either the central galaxy luminosity in units of
$10^{10} \Lsunhh$ or the stellar mass in units of $10^{10} \Msunhh$,
depending upon the property used to stack central galaxies. Clearly,
since the determination of $\sigma_{200}$ requires a sample of
centrals and satellites, this selection method has to be iterative.
Fixed values of the selection criteria parameters are used to identify
the central and the satellite galaxies in the first iteration. The
velocity dispersion of the selected satellites as a function of the
central galaxy property, parameterized via Eq.~(\ref{eq:sig200}), is
fit using a maximum likelihood method and subsequently used to scale
the values of the parameters that define the selection criteria. These
are used to select a new sample of centrals and satellites, and the
entire procedure is repeated until convergence\footnote{We refer the
reader to Paper~II and \citet{vdb04} for details regarding this
method.}.  Using detailed mock galaxy redshift surveys, \citet{vdb04}
have shown that this iterative technique yields much lower interloper
fractions than the more common method using fixed cylindrical volumes
(see also Paper~II). For completeness, and to allow the reader to
reproduce our results, Table~1 lists the final iteration criteria used
for our various samples (in terms of the parameters $a$, $b$, and $c$
that appear in Eq.~[\ref{eq:sig200}]), as well as the total number of
centrals and satellites selected in each sample. Note that these
parameters differ depending on the centrals we choose to stack for our
analysis. The sample of centrals and satellites selected when the
selection criteria are tuned based on the velocity dispersion around
all centrals stacked by luminosity (stellar mass) is called Sample LA
(SA). Samples LR (SR) and LB (SB) are selected by tuning the selection
criteria parameters based upon the velocity dispersion around red and
blue centrals stacked by luminosity (stellar mass), respectively.
\begin{table}
\caption{Selection criteria parameters}
\begin{tabular}{cccccc}
\hline
Samples  & a & b & c & Centrals & Satellites  \\
\hline \hline
    LA   & 2.19  &   0.37   &   0.30   & 3949 & 6213 \\
    LR   & 2.23  &   0.34   &   0.31   & 2723 & 4873 \\
    LB   & 2.11  &   0.47   &  -0.08   & 1082 & 1255 \\
    SA   & 2.07  &   0.22   &   0.20   & 3834 & 6232 \\
    SR   & 2.11  &   0.19   &   0.20   & 3095 & 5445 \\
    SB   & 1.99  &   0.49   &  -0.20   & $\,$701 & $\,$786 \\

\hline
\end{tabular}
\medskip
\begin{minipage}{\hssize}
  The parameters $a,b$ and $c$ that define the criteria used to select
  central and satellite galaxies for all the samples used in this
  paper (see text), and the total numbers of centrals and satellites
  thus selected.
\end{minipage}
\label{tab:table1}
\end{table}

\subsection{Velocity dispersion measurement}
\label{sec:veldisp}

In Paper~I we demonstrated that the commonly measured velocity
dispersion of satellite galaxies, $\sigma_{\rm sat}$, cannot be used
to uniquely determine the scaling relation between halo mass and a
central galaxy property {\it unless this relation has zero scatter}.
In fact, we have shown that different scaling relations with different
amounts of scatter can yield exactly the same $\sigma_{\rm sat}$.  In
the same paper, however, we have shown that this degeneracy can be
broken using a combination of two different measures for the velocity
dispersion of the satellite galaxies: satellite-weighted
($\sigma^2_{\rm sw}$) and host-weighted ($\sigma^2_{\rm hw}$).  To
measure the velocity dispersion in these two different schemes, the
satellite galaxies in the final sample are first binned into
sub-samples based upon the properties (luminosity or stellar mass) of
their central galaxies. For each bin, the distribution of los
velocities of satellite galaxies with respect to their centrals,
$\pdv$, is constructed by either giving each satellite equal weight
(satellite-weighting) or a weight equal to $1/N_{\rm sat}$
(host-weighting), where $N_{\rm sat}$ denotes the number of satellites
around the host of the satellite under consideration. As shown in
Paper~I, the difference between $\sigma^2_{\rm sw}$ and $\sigma^2_{\rm
  hw}$ depends on the amount of scatter in the scaling relation
between halo mass and central galaxy property\footnote{In the case of zero
scatter, one has that $\sigma_{\rm sw} = \sigma_{\rm hw}$.}, and allows the
degeneracy to be broken.

In order to extract the satellite velocity dispersion 
(satellite-weighted or host-weighted) from the corresponding $\pdv$
distributions, we fit $\pdv$ using the sum of two Gaussians plus a
constant:
\begin{equation}
\pdv = a_0 + a_1\,\exp\left[ \frac{-(\dv^2)}{2\sigma_1^2} \right] +
a_2\,\exp\left[ \frac{-(\dv^2)}{2\sigma_2^2} \right]\,.
\end{equation}
The satellite velocity dispersion (satellite-weighted or host-weighted)
then follows from
\begin{equation}
\sigma_{\rm (sw/hw)}^2 = 
\left[\frac{a_1\,\sigma_1^3 + a_2\,\sigma_2^3}{a_1\,\sigma_1 +
a_2\,\sigma_2}\right] - \sigma_{\rm err}^2\,.
\end{equation}
Here $\sigma_{\rm err}$ is the contribution to the effective variance
of $\pdv$ due to redshift errors in the SDSS. Given that each
individual galaxy has a redshift error of $\sim 35 \kms$, the error on
the velocity difference, $\dv$, of the central and satellite galaxies
is $\sigma_{\rm err} = \sqrt{2}\,\times\,35 \simeq 49.5 \kms$, which
is the value we adopt throughout. The errorbars on the velocity
dispersions were estimated as the variance in the velocity dispersions
obtained by fitting $1000$ realisations of the $P(\Delta V)$ histogram
(generated assuming Poisson errorbars) with the procedure described
above.  Detailed tests using mock catalogues have shown that the above
method yields extremely reliable estimates of the actual velocity
dispersions (see Paper II).

Applying this method to the central-satellite samples selected using
the criteria described in Section~\ref{sec:selncrit}, we obtained the
satellite-weighted and host-weighted velocity dispersions as well as
the average number of satellites per central as a function of
luminosity for samples LA, LR and LB, and as a function of stellar
mass for samples SA, SR and SB. In addition, we also measured the
fraction of red central galaxies as a function of luminosity from
Sample LA and as a function of stellar mass from Sample SA. In the
following subsection we describe how to use these data 
to constrain the MLR and MSR of central galaxies.

\subsection{The Model}
\label{sec:model}

The satellite-weighted and host-weighted velocity dispersions of
satellites and the average number of satellites of a central galaxy
with a given property depend on the distribution of halo masses of
central galaxies with that property, $\pmpc$. Therefore these
observables can be used to infer the mean and the scatter of the
scaling relation between halo mass and the central galaxy
property. The analytical expressions that describe these three
quantities are given by (see Paper I):
\begin{equation}
\sigsw^2(\pc) = \frac{ \int_{0}^{\infty} \, \pmpc \, \avnsat_{{\rm
      ap},M} \, \avgsigsatsq_{{\rm ap},M} \, \drm M }
{ \int_{0}^\infty \, \pmpc \, \avnsat_{{\rm ap},M} \drm M } \,,
\label{sweqnvol} 
\end{equation}
\begin{equation}
\sighw^2(\pc) =
\frac{ \int_{0}^{\infty} \, \pmpc \, {\cal P}(M) \,
\avgsigsatsq_{{\rm ap},M} \, \drm M} { \int_{0}^\infty \, \pmpc \,
{\cal P}(M) \, \drm M }\,,
\label{hweqnvol}
\end{equation}
\begin{equation}
\avnsat(\pc) =
\frac{ \int_{0}^{\infty} \pmpc \avnsat_{{\rm ap},M} \drm M} 
{ \int_{0}^\infty \pmpc  {\cal P}(M) \drm M }\,.
\label{nsateqnvol}
\end{equation}
Here $\avnsat_{{\rm ap},M}$ and $\avgsigsatsq_{{\rm ap},M}$ denote the
average number of satellites and the average velocity dispersion of
satellites within the aperture $R_{\rm s}$ in a halo of mass $M$,
respectively. The factor ${\cal P}(M)$ is the fraction of haloes of
mass $M$ that host at least one satellite. 
Our way of modelling the
observables thus consists of two parts: (i) the kinematics of
satellite galaxies in a single halo of a given mass and (ii) the
statistics that describe how central and satellite galaxies occupy
haloes. We describe each of these parts in the following subsections.

\subsubsection{Kinematics in a single halo}

We assume that dark matter haloes are spherically symmetric and that
their density distributions follow the universal NFW profile
\citep{Navarro1997},
\begin{equation}\label{rhor}
 \rho(r|M) \propto \left( \frac{r}{r_s}\right)^{-1} \left( 1 +
 \frac{r}{r_s} \right)^{-2}\,,
\end{equation}
where $r_s$ is a scale radius specified in terms of the virial radius
$r_{\rm vir}$ using the concentration parameter, $c=r_{\rm
  vir}/r_s$. We use the concentration-mass relation from
\cite{Maccio2007}, appropriately modified for our definition of the
halo mass. We assume that the number density distribution of satellite
galaxies, $n_{\rm sat}(r|M)$, is given by
\begin{equation}\label{nsr}
n_{\rm sat}(r|M) \propto \left(
\frac{r}{ {\cal R} r_s}\right)^{-\gamma}
\left( 1 + \frac{r}{ {\cal R} r_s} \right)^{\gamma-3}.
\end{equation}
Here $\gamma$ represents the slope of the number density distribution
of satellites as $r \rightarrow 0$ and ${\cal R}$ is a free parameter.
Throughout this paper, we use the result from Paper II that the number
density distribution of satellites in the SDSS can be well described
by Eq.~(\ref{nsr}) with $\gamma=0.0$ and ${\cal R}=2$ (see also
\citealt{Yang2005} and \citealt{More2009c}). The distribution $n_{\rm
  sat}(r|M)$ is normalized such that
\begin{equation}
 \avnsatm = 4 \pi \, \int_0^{r_{\rm vir}} n_{\rm sat}(r|M) \, r^2 \,
 \drm r \,,
\end{equation}
where $\avnsatm$ denotes the average number of satellites in a halo of mass
$M$. The number of satellites within the aperture, $\avnsat_{{\rm ap},M}$, is
then given by
\begin{equation}\label{nsatap}
 \avnsat_{{\rm ap},M} = 4 \pi \int_0^{R_{\rm s}} R \, \rmd R
\int_{R}^{r_{\rm vir}} n_{\rm sat}(r|M) \, {r \rmd r \over \sqrt{r^2
- R^2}}\,.
\end{equation}
We further assume that the satellite occupation numbers follow Poisson
statistics, which is supported both by direct observations
\citep[e.g.][]{Yang2008} and by numerical simulations
\citep{Kravtsov2004}. The fraction of central galaxies that have at
least one satellite within the aperture radius $R_{\rm s}$ is then
given by
\begin{eqnarray}
{\cal P}(M) = 1 - \exp[-\avnsat_{{\rm ap},M}].
\label{poisson}
\end{eqnarray}

To find an analytical expression for $\avgsigsatsq_{{\rm ap},M}$,
first note that the Jeans equation can be used to find an expression
for the radial velocity dispersion of satellites at a radial distance
$r$ from the centre;
\begin{eqnarray}
\sigma^2_{\rm sat}(r|M) =  {c \, V^2_{\rm vir} \over {\cal R}^2
\mu(c)} \, \left({r \over {\cal R} r_s}\right)^\gamma \,
\left(1 + {r \over {\cal R} r_s}\right)^{3-\gamma} \, \nonumber \\
\int_{r/r_s}^{\infty} {\mu(x) {\rm d}x  \over
~(x / {\cal R})^{\gamma + 2}(1+ x / {\cal R} )^{3-\gamma}\,}\,.
\label{sig1dm}
\end{eqnarray}
Here $V_{\rm vir}$ is the circular velocity at $r_{\rm vir}$ and
\begin{equation}
\mu(x) = \int_0^x y (1+y)^{-2} \rmd y\,, 
\end{equation}
(see Paper I for a detailed derivation).  If the velocity dispersion
of the satellites is assumed to be isotropic, then $\avgsigsatsq_{{\rm
ap},M}$ can be expressed as the average of the radial velocity
dispersion, $\sigma_{\rm sat}^2(r|M)$, over the aperture $R_{\rm s}$:
\begin{eqnarray}\label{sigaper}
\lefteqn{\avg{\sigma_{\rm sat}^2}_{{\rm ap},M} = { 4 \pi \over \avnsat_{{\rm
ap},M}} \int_0^{R_{\rm s}} R \, {\rm d}R} \nonumber \\
& & \int_{R}^{r_{\rm vir}} n_{\rm sat}(r|M) \, \sigma_{\rm
sat}^2(r|M) \, {r \rmd r\over \sqrt{r^2 - R^2}}\,.
\end{eqnarray}

\subsubsection{Halo occupation statistics}
\label{sec:hos}

To model the kinematics of satellite galaxies around central galaxies
stacked by a particular property, we need to specify the distribution
of halo masses for central galaxies with that property, $\pmpc$. Let
us first consider the case of central galaxies stacked by luminosity
and colour. In this case, $\pc\equiv\lc \cap C$, where $L$ denotes the
luminosity of the central galaxy and $C$ its colour. We will use the
letters R for red and B for blue when referring to the actual
colours. Using Bayes' theorem,
\begin{equation}
\label{bayes1}
P(M| \lc\cap C) %&=& \frac{ P(M \cap \lc\cap C) }{ P(\lc\cap C) } \nonumber \\
                %&=& \frac{ P(\lc| M \cap C)\,P(M \cap C) }{ P(\lc\cap C) } \nonumber \\
                = \frac{ P(\lc| M \cap C)\,f_{C}(M)\,P(M) }{f_{C}(\lc)\,P(\lc) }
\end{equation}
Here $P(\lc| M \cap C)$ describes the distribution of central galaxy
luminosities in haloes of mass $M$ that host central galaxies of a
particular colour $C$, $f_{C}(M) = P(C|M)$ is the fraction of haloes
of mass $M$ that host a central galaxy of colour $C$, and $P(M)$ is
proportional to the halo mass function $n(M)$\footnote{For the
  analysis in this paper, we use the halo mass function of
  \citet{Tinker2008} for which haloes are defined as spheres with an
  average density that is 200 times the average matter density in the
  universe.}. Note that the denominator has no dependence on $M$ and
is just a multiplicative normalisation constant which cancels out when
we model our observables (see Eqs.~\ref{sweqnvol}~-~\ref{nsateqnvol}).

We use simple parametric forms to model the distribution $P(\lc| M
\cap C)$ and $f_{C}(M)$. The distribution $P(\lc| M \cap C)$ is
assumed to be a log-normal given by
\begin{equation}\label{lognormal}
 P(\lc| M \cap C) = {\log(e) \over \sqrt{2 \pi} \siglcen}
\exp\left[ - {(\log [\lc/\tilde{\lc}])^2 \over 2 \siglcen^2 } \right]
\frac{\drm \lc}{\lc}\,.
\end{equation}
Here $\log \tilde{\lc} (C,M)$ denotes the mean of the log-normal
distribution and $\siglcen(C)$ is the corresponding scatter. We use
four parameters each to specify the relations $\tilde{\lc}(R,M)$ and
$\tilde{\lc}(B,M)$: a low mass end slope, $\gamma_1$, a high mass end
slope, $\gamma_2$, a characteristic mass scale, $M_1$, and a
normalisation, $L_0$;
\begin{equation}
\label{ltilde}
\tilde{\lc} = L_0  \frac{ \left(M/M_1\right)^{\gamma_1} }{ [1
+ \left( M/M_1\right)]^{\gamma_1-\gamma_2}}.
\end{equation}
We further assume that the scatters $\siglcen(R)$ and $\siglcen(B)$
are independent of halo mass. Thus for each colour C, we use $5$
parameters to describe the distribution $P(\lc| M \cap C)$. This
parametrization is motivated by the results of \citet{Yang2008}, who
inferred the conditional luminosity function using the large SDSS
galaxy group catalogue of \citet{Yang2007}.

The function $f_R(M)$ is assumed to be linear in $\log M$:
\begin{eqnarray}
\label{frm}
f'_R(M) & = & f_{0} + \alpha_f\,[\log (M/\Msunh)-12.0] \nonumber \\
f_R(M)  & = & \min ( \max[ 0, f'_R(M)], 1.0)
\end{eqnarray}
where the second equality takes into account that $f_R$ is a fraction,
and therefore bounded by zero and unity \citep[cf.][]{vdb03}.
Also note that $f_{B}(M)=1-f_R(M)$, as `red' and `blue' form a
mutually exclusive and exhaustive set of colours assigned to the
central galaxies. Hence, $f_R(M)$ and $f_B(M)$ add a total of two free
parameters to our model.

We also need a model for the satellite occupation numbers,
$\avnsatm$. Throughout we assume that the number of satellite
galaxies in a halo of mass $M$ scales with halo mass as
\begin{equation}
\label{nsatm}
\avnsatm = N_{12} \left(\frac{M}{10^{12} \Msunh}\right)^{\alpha}\,,
\end{equation}
which adds two more parameters to our model; $N_{12}$ and $\alpha$.
Note that we assume that $\avnsatm$ is independent of the colour of
the central galaxies. Although we believe this to be a realistic
assumption, we will discuss the potential impact of its violations in
Section~\ref{sec:results}. 

Next, consider the case where galaxies are stacked only according to
their luminosity, i.e., $\pc \equiv \lcen$. To model the observables
we need to know the distribution 
\begin{equation}
P(M|\lc) = \frac{ P(\lc|M)\,P(M) }{ P(\lc) } \propto P(\lc|M)\,n(M)\,.
\end{equation}
The distribution $P(\lc|M)$ is related to the distributions $P(\lc
\cap R|M)$ and $P(\lc \cap B|M)$ according to:
\begin{equation}
P(\lc|M) = P(\lc|M\cap R) f_R(M) + P(\lc|M\cap B) f_B(M)
\end{equation}
Finally, the expression that describes the fraction of red centrals as
a function of central galaxy luminosity is given by 
\begin{equation}
f_R(L) = \frac{\int_{0}^{\infty} P(L|M \cap R) \, {\cal P}(M) \, n(M) \, f_R(M) 
\drm M}{\int_{0}^{\infty} P(\lc|M) \, {\cal P}(M) \, n(M) \drm M }
\end{equation}
Note that we have appropriately corrected for the fact that the
observed fraction of red centrals is calculated using only those
centrals that have at least one satellite. 

Hence, our analytical model has a total of 14 free parameters and
completely describes the kinematics of satellite galaxies and the
average number of satellites around centrals stacked by luminosity in
Samples LA, LR and LB. This analytical framework also allows us to
calculate the fraction of red centrals as a function of luminosity.

For the analysis of the kinematics of satellite galaxies around
centrals stacked by stellar mass, there is an additional complication
that has to be addressed. The central and satellite galaxies used for
our analysis are selected from a volume limited sample that is
complete above a certain luminosity. Since we have used both colour
and luminosity to assign the stellar masses, our sample starts to
become incomplete in stellar mass roughly below $10^{10}\Msunhh$.  The
completeness is a function of both stellar mass and colour, and is
described by the sample selection function $S(M_*,C)$, defined as the
fraction of galaxies of stellar mass $M_*$ and colour $C$ in the SDSS
volume with $0.02 \leq z \leq 0.072$ (i.e., our sample volume) that
make it into the sample. The determination of $S(M_*,C)$ is discussed
in Appendix~\ref{app:a}. 

The sample selection function $S(M_*,C)$ enters our model in the
following way. We can write the stacking property as $\pc \equiv
M_*\cap C \cap \Sset$, where we use $\Sset$ to denote the subset of
all galaxies with stellar mass $M_*$ and colour $C$ in our sample
volume that make it into the sample.  The corresponding distribution
of halo masses, $P(M|M_*\cap C \cap \Sset)$, can then be written as
\begin{eqnarray}
\lefteqn{P(M|M_*\cap C \cap \Sset) = 
\frac{ P(\Sset | M \cap M_* \cap C) \, P(M \cap M_* \cap C)}
{P(M_* \cap C \cap \Sset)}}  \nonumber \\
& & = \frac{ S(M_*,C) } {P(M_* \cap C \cap \Sset)}\, f_C(M) \, P(M) \,
P(M_* | M \cap C)\,.
\end{eqnarray}
In the second equality, we have identified the distribution $P(\Sset|M
\cap M_* \cap C)$ as the selection function, $S(M_*,C)$. In addition,
we have also expressed $P(M \cap M_* \cap C)$ in terms of $P(M_*|M
\cap C)$. Note that the selection function does not depend on the halo
mass $M$ and acts as a multiplicative normalisation constant for the
distribution $P(M|M_*\cap C \cap \Sset)$. It harmlessly cancels out
from the expressions that analytically describe the observables when
central galaxies are stacked by colour. However, it turns out to be
important for calculating $P(M|M_* \cap \Sset)$ and hence the
observables when central galaxies are stacked by stellar mass
alone. First note that the distribution $P(M|M_* \cap \Sset)$ is
related to $P(M_*|M\cap \Sset)$ such that
\begin{equation}
P(M|M_* \cap \Sset)=\frac{P(M_*\cap \Sset|M)\,P(M)}{P(M_* \cap \Sset)}\,.
\end{equation}
The distribution $P(M_*\cap \Sset | M)$ can be expressed in terms of
$P(M_*|M\cap R)$ and $P(M_*|M \cap B)$ as follows:
\begin{eqnarray}
\lefteqn{P(M_*\cap \Sset | M) = P(M_*\cap R \cap \Sset|M)
+ P(M_*\cap B \cap \Sset|M)} \nonumber \\
& & = \, f_R(M) \, S(M_*,R) \, P(M_*|M\cap R) + \nonumber \\
& & \,\,\,\,\; f_B(M) \, S(M_*,B) \, P(M_*|M\cap B)
\end{eqnarray}

Similar to Eq.~(\ref{lognormal}), we parameterize $P(M_*|M \cap C)$ as
a log-normal distribution with mean $\log \tilde{M_*}(C,M)$ and a
scatter $\sigmscen$ which depends on colour but is independent of halo
mass. The relation $\tilde{M_*}(C,M)$ is described using
Eq.~(\ref{ltilde}), but with $L$ replaced by $M_*$, and we assume that
the scatters $\sigma_{\log M_*}(R)$ and $\sigma_{\log M_*}(B)$ are
independent of halo mass. For $f_R(M)$ and $\avnsatm$ we adopt the
same parameterizations as before (i.e., Eqs.~[\ref{frm}]
and~[\ref{nsatm}]).  Hence, our model for the analysis of the MSR also
contains 14 free parameters, which we constrain using the observables
obtained by stacking central galaxies by stellar mass and colour.

\subsection{Constraining the model parameters} 

We now describe our method to constrain the model parameters.  Here we
focus on the analysis of the MLR, but note that the analysis of the
MSR is basically the same. We have measurements of the
satellite-weighted velocity dispersion, the host-weighted velocity
dispersion and the average number of satellites per central, for 10
different luminosity bins, and for each of the three samples LA, LR
and LB.  In addition, we have 10 measurements of the fraction of red
centrals as a function of luminosity. Since most of the centrals in
sample LA are present in either sample LR or sample LB, the velocity
dispersions and average number of satellites measured from sample LA
are not independent from those obtained using samples LR and LB.
Therefore, we do not use these measurements from sample LA to
constrain the model parameters. This leaves a total of 70 independent
data points to constrain our 14 model parameters.

We use flat uninformative priors on each of the model parameters
(albeit in a limited interval for each of the parameters). We use a
Monte-Carlo Markov Chain (hereafter MCMC) technique to sample from the
posterior probability distribution of each of these parameters given
the observational constraints.  The MCMC is a chain of models, each
consisting of the 14 parameters. At any point in the chain, a trial
model is generated with the 14 free parameters drawn from a
14-dimensional Gaussian proposal distribution which is centered on the
current values of the parameters. The chi-squared statistic,
$\chi^2_{\rm try}$, for this trial model, is calculated using
\begin{equation}
 \chi^2_{\rm try} = \sum_{C=R,B} \left[ \chi^2_{\rm sw}(C) + \chi^2_{\rm hw}(C) 
                  + \chi^2_{\rm ns}(C) \right] + \chi^2_{\rm fr}\, ,
\end{equation}
where
\begin{eqnarray}
 \chi^2_{\rm sw}(C) &=& \sum_{i=1}^{10} \left[{\sigsw(\pc[i]) -
\hat{\sigma}_{\rm sw}(\pc[i]) \over \Delta \hat{\sigma}_{\rm
sw}(\pc[i])} \right]^2 \,, \\
\chi^2_{\rm hw}(C) &=& \sum_{i=1}^{10} \left[{\sighw(\pc[i])
- \hat{\sigma}_{\rm hw}(\pc[i]) \over \Delta \hat{\sigma}_{\rm
hw}(\pc[i])} \right]^2 \, , \\
\chi^2_{\rm ns}(C) &=& \sum_{i=1}^{10} \left[{\avnsat(\pc[i])
- {\hat{N}_{\rm s}}(\pc[i]) \over \Delta {\hat{N}_{\rm
s}}(\pc[i])}
\right]^2 \,, \\
\chi^2_{\rm fr} &=& \sum_{i=1}^{10} \left[{f_{R}(\lc)
- {\hat{f}_R}(\lc[i]) \over \Delta {\hat{f}_R}(\lc[i])}
\right]^2 \,.
\end{eqnarray}
Here $\pc\equiv L\cap C$, $\hat{X}$ denotes the observable $X$ and
$\Delta \hat{X}$ its corresponding error. The trial step is accepted
with a probability given by
\begin{equation}
P_{\rm accept}=\left\{ \begin{array}{cl}
        1.0, & {\rm if}\; \chi^2_{\rm try} \le \chi^2_{\rm cur} \\
        {\rm exp}[-(\chi^2_{\rm try}-\chi^2_{\rm cur})/2], & {\rm if}
\; \chi^2_{\rm try} > \chi^2_{\rm cur}
           \end{array}\right.
\end{equation}
where $\chi^2_{\rm cur}$ denotes the $\chi^2$ for the current model in
the chain.

We initialize the chain from a random position in our 14-dimensional
parameter space and discard the first $20,000$ models (the `burn-in'
period) allowing the chain to sample from a more probable part of the
distribution.  We proceed and construct a chain consisting of 40
million models.  We thin this chain by a factor of $10^3$ to remove
the correlations between neighbouring models. This leaves us with a
chain of $40,000$ independent models that sample the posterior
distribution. We use this chain of models to estimate the confidence
levels on the parameters and the relations of interest, namely the
mean and the scatter of the scaling relation between halo mass and the
central galaxy property under consideration.

\begin{table}
\caption{MLR: Percentiles of the posterior distributions}
\begin{tabular}{ccccc}
\hline
         & Parameter  & 16 percent  & 50 percent & 84 percent  \\
\hline
\hline
Red       & $\log(L_0)$   &  9.62  &  9.87 & 10.24  \\
centrals  & $\log(M_1)$   & 11.20  & 11.55 & 11.96  \\
          & $\gamma_1 $   &  2.42  &  3.37 &  4.43  \\
          & $\gamma_2 $   &  0.26  &  0.36 &  0.43  \\
          & $\siglcen$    &  0.18  &  0.20 &  0.23  \\
\hline
Blue      & $\log(L_0)$   &  8.98  &  9.32 &  9.67  \\
centrals  & $\log(M_1)$   & 10.14  & 10.45 & 10.91  \\
          & $\gamma_1 $   &  2.29  &  3.08 &  4.26  \\
          & $\gamma_2 $   &  0.32  &  0.46 &  0.58  \\
          & $\siglcen$    &  0.13  &  0.20 &  0.26  \\
\hline
All       & $f_0$         &  0.57  &  0.64 &  0.71  \\
centrals  & $\alpha_f$    &  0.14  &  0.20 &  0.27  \\
          & $\log N_{12}$ & -0.83  & -0.72 & -0.62  \\
          & $\alpha$      &  1.19  &  1.28 &  1.38  \\
\hline
\end{tabular}
\medskip
\begin{minipage}{\hssize}
  The 16, 50 and 84 percentile values of the posterior distributions
  for the parameters of our model obtained from the MCMC analysis of
      the velocity dispersion data from Samples LR, LB and LA.
\end{minipage}
\label{tab:table2}
\end{table}
\begin{table}
\caption{MSR: Percentiles of the posterior distribution}
\begin{tabular}{ccccc}
\hline
         & Parameter  & 16 percent  & 50 percent & 84 percent  \\
\hline
\hline
Red      & $\log(M_{*0})$ & 10.12  & 10.63  & 10.97  \\
centrals & $\log(M_1)$    & 11.42  & 11.94  & 12.29  \\
         & $\gamma_1 $    &  2.44  &  3.44  &  4.49  \\
         & $\gamma_2 $    &  0.15  &  0.29  &  0.44  \\
         & $\sigcenm$     &  0.16  &  0.19  &  0.22  \\
\hline
Blue     & $\log(M_{*0})$ &  8.36  &  9.42  & 10.68  \\
centrals & $\log(M_1)$    & 10.57  & 11.29  & 11.98  \\
         & $\gamma_1 $    &  2.32  &  3.21  &  4.37  \\
         & $\gamma_2 $    &  0.48  &  0.98  &  1.31  \\
         & $\sigcenm$     &  0.08  &  0.15  &  0.27  \\
\hline
All       & $f_0$         &  0.37  &  0.45 &  0.56  \\
centrals  & $\alpha_f$    &  0.43  &  0.61 &  0.77  \\
          & $\log N_{12}$ & -0.91  & -0.79 & -0.68  \\
          & $\alpha$      &  1.18  &  1.27 &  1.38  \\
\hline

\end{tabular}
\medskip
\begin{minipage}{\hssize}
  The 16, 50 and 84 percentile values
  of the posterior distributions for the parameters of our model
  obtained from the MCMC analysis of the velocity dispersion data from
  Samples SR, SB and SA.
\end{minipage}
\label{tab:table3}
\end{table}

\section{Results} 
\label{sec:results} 

\begin{figure*}
\centerline{\psfig{figure=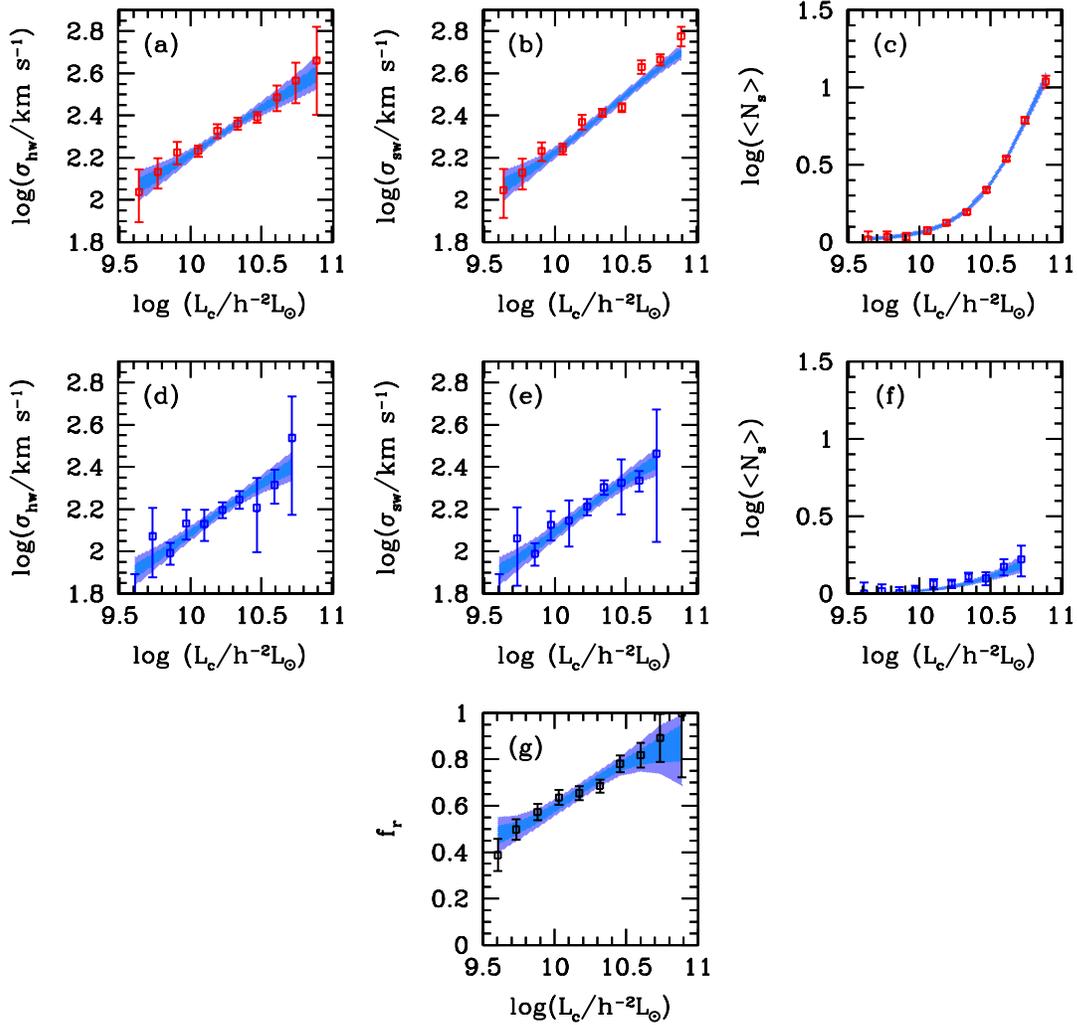,width=0.8\hdsize}}
\caption{Observables used to constrain the MLR of central galaxies
  (open squares with errorbars). The upper and middle panels show the
  observables measured using red centrals (Sample LR) and blue
  centrals (Sample LB), respectively. From the left to the right these
  panels show the host-weighted velocity dispersions [panels (a) and
  (d)], the satellite-weighted velocity dispersions [panels (b) and
  (e)], and the average number of satellites per central [panels (c)
  and (f)], all as function of the luminosity of the central. Panel
  (g) shows the fraction of red centrals as a function of luminosity
  as measured from Sample LA.  The blue and purple regions indicate
  the 68 and 95 percent confidence intervals obtained from the MCMC,
  showing that the model accurately fits the data.
  }
\label{fig:fig1_MLR}
\end{figure*}
\begin{figure*}
\centerline{\psfig{figure=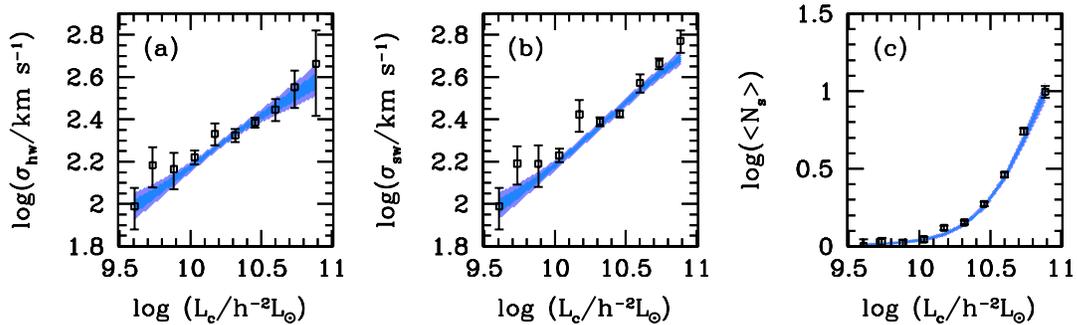,width=0.8\hdsize}}
\caption{Obervables measured using Sample LA. Panels (a), (b) and (c)
  shows the host-weighted velocity dispersions, the satellite-weighted
  velocity dispersions, and the average number of satellites per
  central, respectively, all as a function of the luminosity of the
  central galaxy. Although these observables are not used to constrain
  the model parameters (see text), the 68 and 95 percent confidence
  intervals obtained from the MCMC, indicated by the blue and purple
  regions, show that the model accurately fits these data as well.}
\label{fig:fig2_MLR}
\end{figure*}
\begin{figure*}
\centerline{\psfig{figure=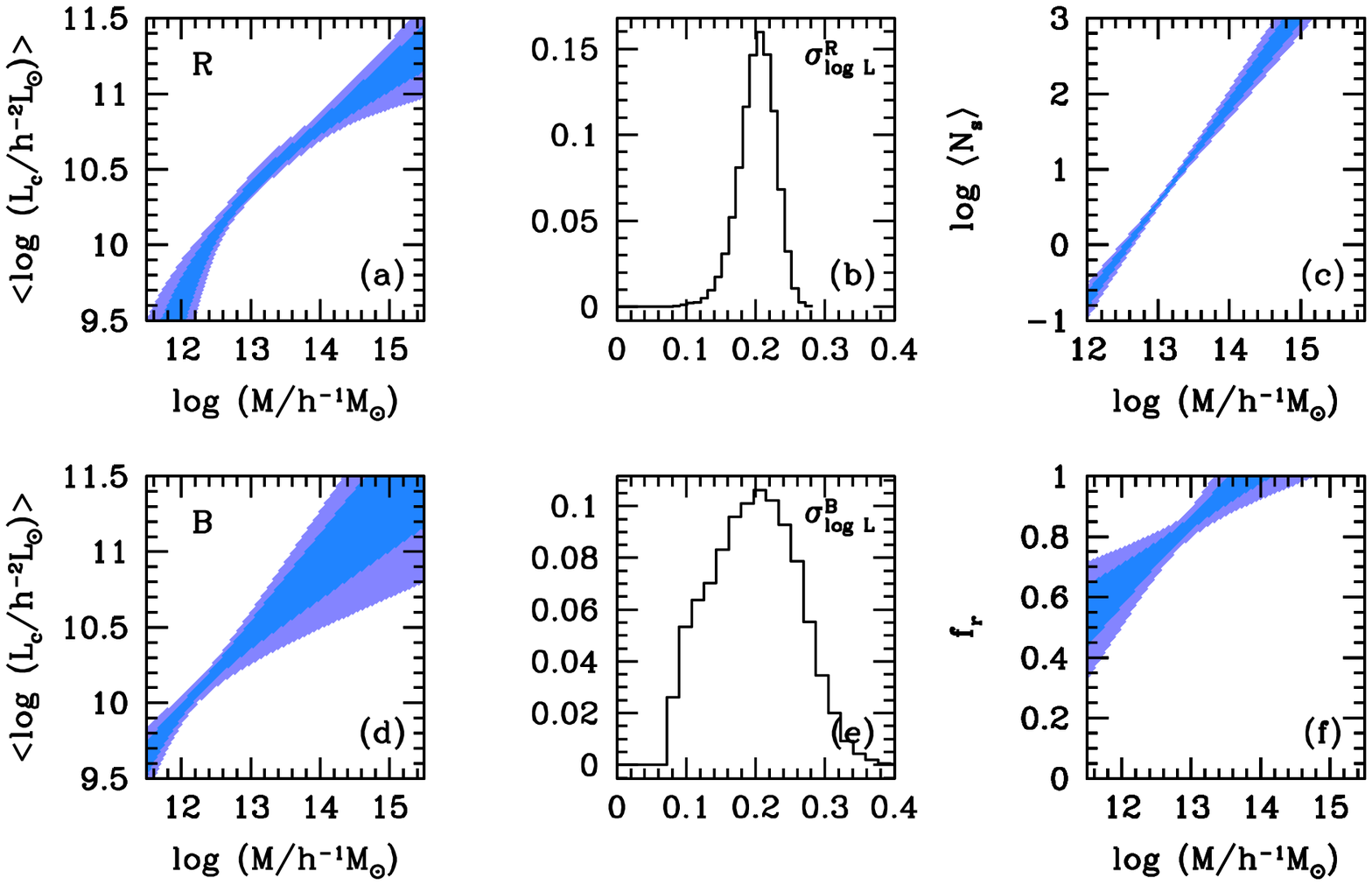,width=0.8\hdsize}}
\caption{Constraints on the model ingredients obtained from the MCMC.
  The 68 and 95 percent confidence intervals are indicated as blue and
  purple regions, respectively. Panel (a) shows the average luminosity
  of red centrals as a function of halo mass, while panel (b) shows
  the posterior distribution of the scatter $\siglcen$ in this
  relation.  Panels (d) and (e) show the same but for blue
  centrals. Panel (c) shows the constraints on the average number of
  satellites as a function of halo mass, and finally, panel (f) shows
  the constraints on the fraction of red centrals as a function of
  halo mass. 
  }
\label{fig:fig3_MLR}
\end{figure*}
\begin{figure*}
\centerline{\psfig{figure=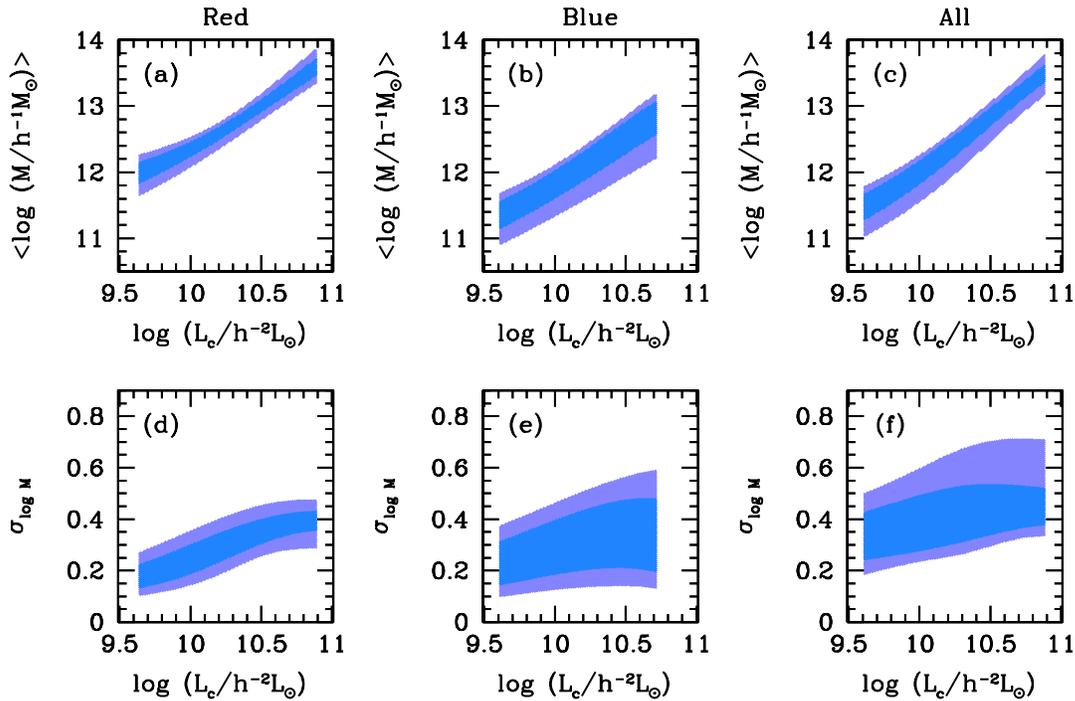,width=0.8\hdsize}}
\caption{Constraints on the MLR obtained from the MCMC. The average
  halo masses as a function of the luminosity of red centrals, blue
  centrals, and all centrals are shown in panels (a), (b) and (c)
  respectively. Panels (d), (e) and (f) show the confidence intervals
  on the scatter in these relations. As in the previous figures, blue
  and purple colours indicate the 68 and 95 percent confidence
  intervals. 
  }
\label{fig:fig4_MLR}
\end{figure*}
\begin{figure*}
\centerline{\psfig{figure=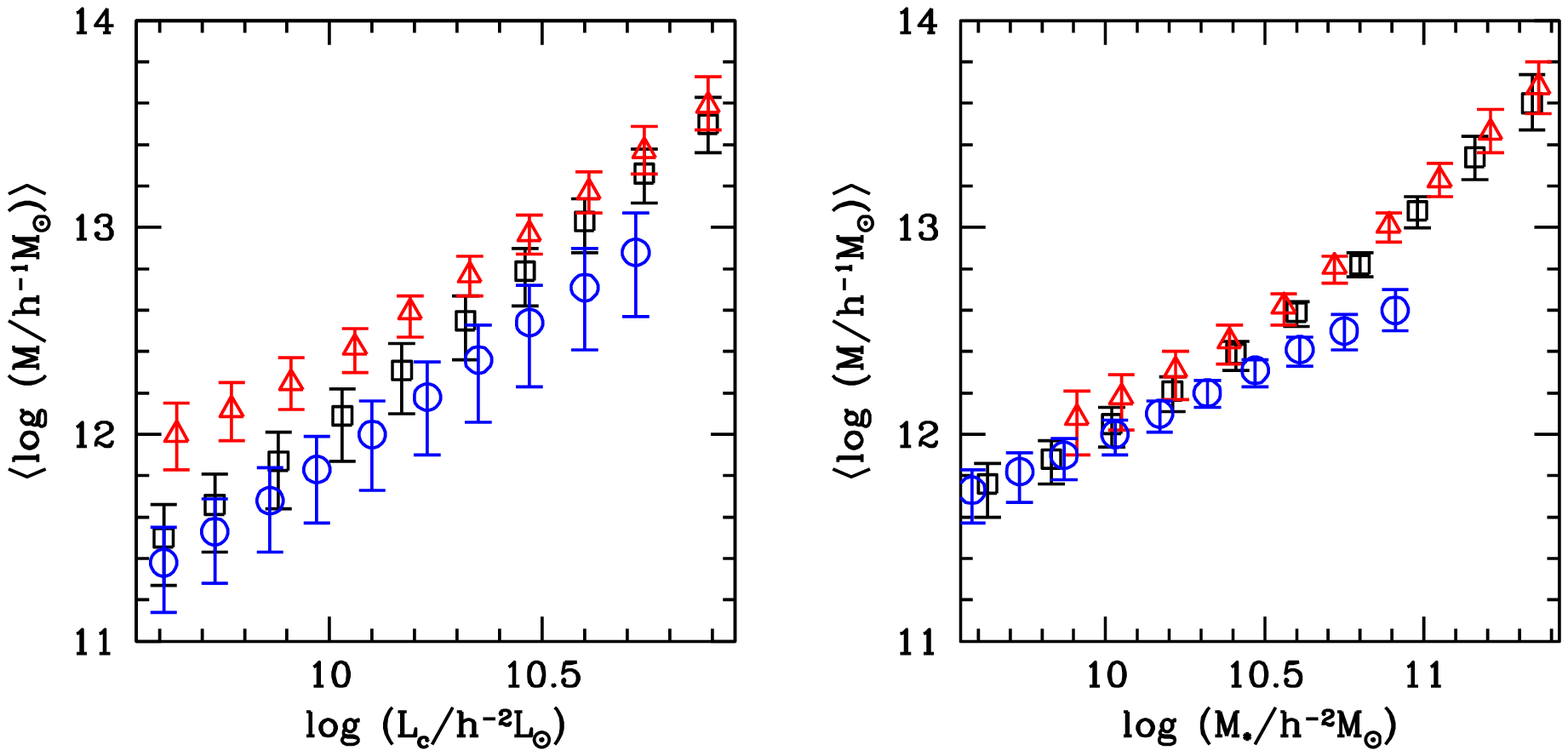,width=0.8\hdsize}}
\caption{Comparison between the average MLR (left hand panel) and the
  average MSR (right hand panel) of all centrals (squares), red
  centrals (triangles) and blue centrals (circles). Errorbars indicate
  the $68$ percent confidence intervals. See text for discussion.
  }
\label{fig:fig5_MLRMSR}
\end{figure*}
\subsection{The Halo Mass$-$Luminosity Relation}

The analysis of the MLR of central galaxies is carried out by
analyzing Samples LA, LR and LB (see Table~1 for the selection
criteria and the numbers of centrals and satellites in each of these
samples).  The host-weighted velocity dispersion, the
satellite-weighted velocity dispersion and the average number of
satellites as a function of the luminosity of the central galaxies
obtained from Samples LR are shown as open squares in panels (a), (b)
and (c) of Fig.~\ref{fig:fig1_MLR} respectively. In the same figure,
panels (d), (e) and (f) show the corresponding observables obtained
from Sample LB. At fixed luminosity, the velocity dispersion of
satellite galaxies around red centrals are systematically larger than
the velocity dispersions around blue centrals. The same is also true
for the average number of satellites. The fraction of red centrals
as a function of luminosity obtained from Sample LA are shown in panel
(g). The data in these 7 panels is used to constrain the 14 parameters
of our model that describe the halo occupation statistics of red and
blue centrals, and the satellite occupation numbers in haloes that
host red and blue centrals. The blue and purple shaded regions
indicate the 68 and 95 percent confidence levels obtained from our
MCMC. A comparison with the data reveals that the model accurately
reproduces the data.

Fig.~\ref{fig:fig2_MLR} shows the velocity dispersions and the average
number of satellites as a function of luminosity around {\it all}
centrals obtained from Sample LA. As for samples LR and LB, the model
is in excellent agreement with the data (open squares), even though
these data were not directly used to constrain the model.

The 16, 50 and 84 percentile values of the posterior distribution for
our model parameters obtained from our analysis are listed in
Table~\ref{tab:table2}. The constraints on our model ingredients are
presented pictorially in Fig.~\ref{fig:fig3_MLR}.  Panels (a) and (d)
show the constraints obtained on $\log \tilde{\lc} (M)$ for red and
blue centrals, respectively, while panels (b) and (e) show the
corresponding posterior distributions of the scatter in luminosities
at fixed halo masses. Panel (c) shows the average number of satellite
galaxies as a function of halo mass, and panel (f) shows the fraction
of red centrals as a function of halo mass.  At the bright end, the
mean luminosity of red central galaxies scales with halo mass as $\lc
\propto M^{ 0.36 ^{+0.07} _{-0.10} }$ while that of blue central
galaxies scales as $\lc \propto M^{0.46^{+0.12}_{-0.14}}$. At the
faint end, the constraints on the faint end slope of the $\log
\tilde{\lc} (M)$ relation are entirely dominated by the prior
$\gamma_1 \in [2.0,5.0]$. This is due to the magnitude limit of the
SDSS, which is not sufficiently faint to reliably probe the occupation
statistics of dark matter haloes with $M \lta 10^{12} h^{-1} \Msun$
\citep[but see][]{Yang2009}.  For the scatter in $\plcm$ we obtain
that $\siglcen= 0.20^{+0.03}_{-0.02}$ for red centrals and $\siglcen=
0.20^{+0.06}_{-0.07}$ for blue centrals.  Note that the scatter for
blue centrals is less well contrained than for red centrals, which is
due to the smaller sample size (see Table~1).

Fig.~\ref{fig:fig4_MLR} shows the average halo mass and the scatter in
halo masses obtained from our analysis for red centrals (left-hand
panels), blue centrals (middle panels) and all centrals (right-hand
panels) as a function of luminosity. As expected, brighter galaxies
reside on average in more massive haloes. Note that the scatter in
halo masses around all centrals appears somewhat higher than that
around either red or blue centrals. This indicates that some fraction
of this scatter is due to the fact that red and blue centrals of the
same luminosity reside, on average, in haloes of different mass (see
below). However, red and blue centrals separately still reveal a
significant amount of scatter in their halo masses, whereby the
scatter around red centrals increases with luminosity, while that
around blue centrals shows no clear luminosity dependence.  

The left-hand panel of Fig.~\ref{fig:fig5_MLRMSR} summarizes these
results. It compares the MLR of red centrals (triangles) to that of
blue centrals (circles) and to that of all centrals (squares). At
fixed luminosity, red centrals on average reside in more massive
haloes than blue centrals. As expected, the MLR of all centrals lies
in between that of red and blue centrals. As the fraction of red
centrals steadily increases to unity at the bright end, the average
halo mass of all centrals shifts from tracing the MLR of blue centrals
to tracing the MLR of red centrals.

\subsection{The Halo Mass$-$Stellar Mass Relation} 
\begin{figure*}
\centerline{\psfig{figure=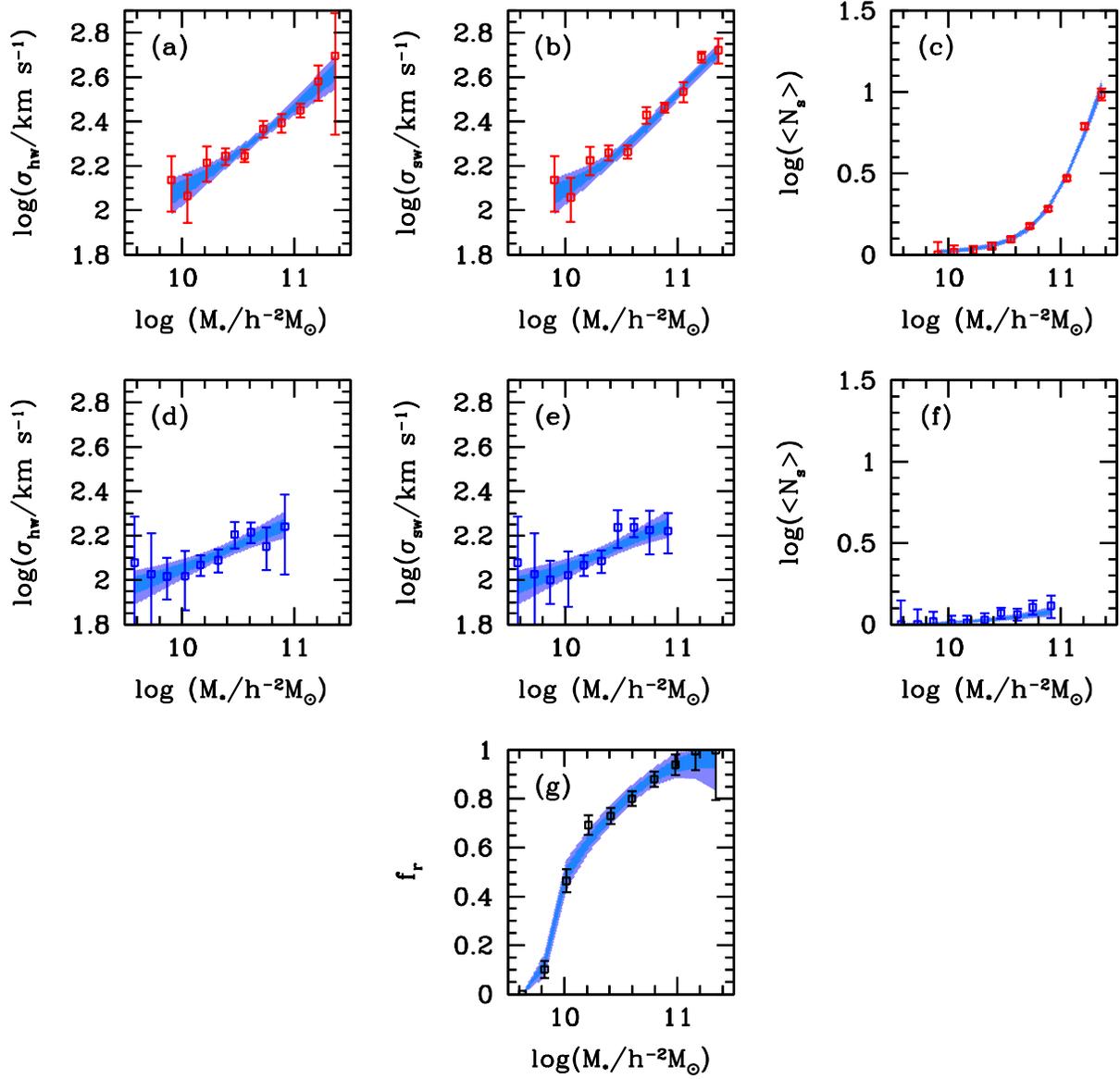,width=0.9\hdsize}}
\caption{Same as Fig.~\ref{fig:fig1_MLR}, but as function of the
  stellar mass of the centrals, based on samples SR, SB and SA.}
\label{fig:fig1_MSR}
\end{figure*}
\begin{figure*}
\centerline{\psfig{figure=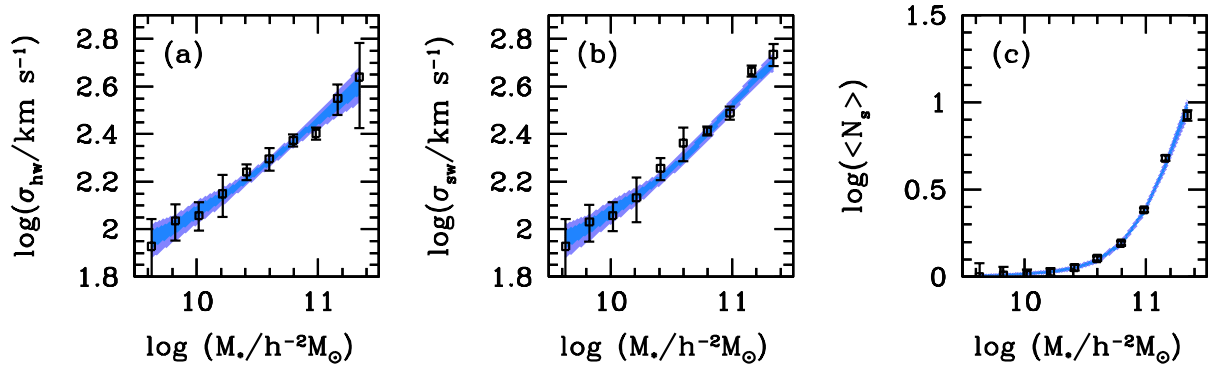,width=0.9\hdsize}}
\caption{Same as Fig.~\ref{fig:fig2_MLR}, but as function of the
  stellar mass of the centrals, based on sample SA.  As for
  Fig.~\ref{fig:fig2_MLR}, these observables are not used to constrain
  the model parameters.}
\label{fig:fig2_MSR}
\end{figure*}
\begin{figure*}
\centerline{\psfig{figure=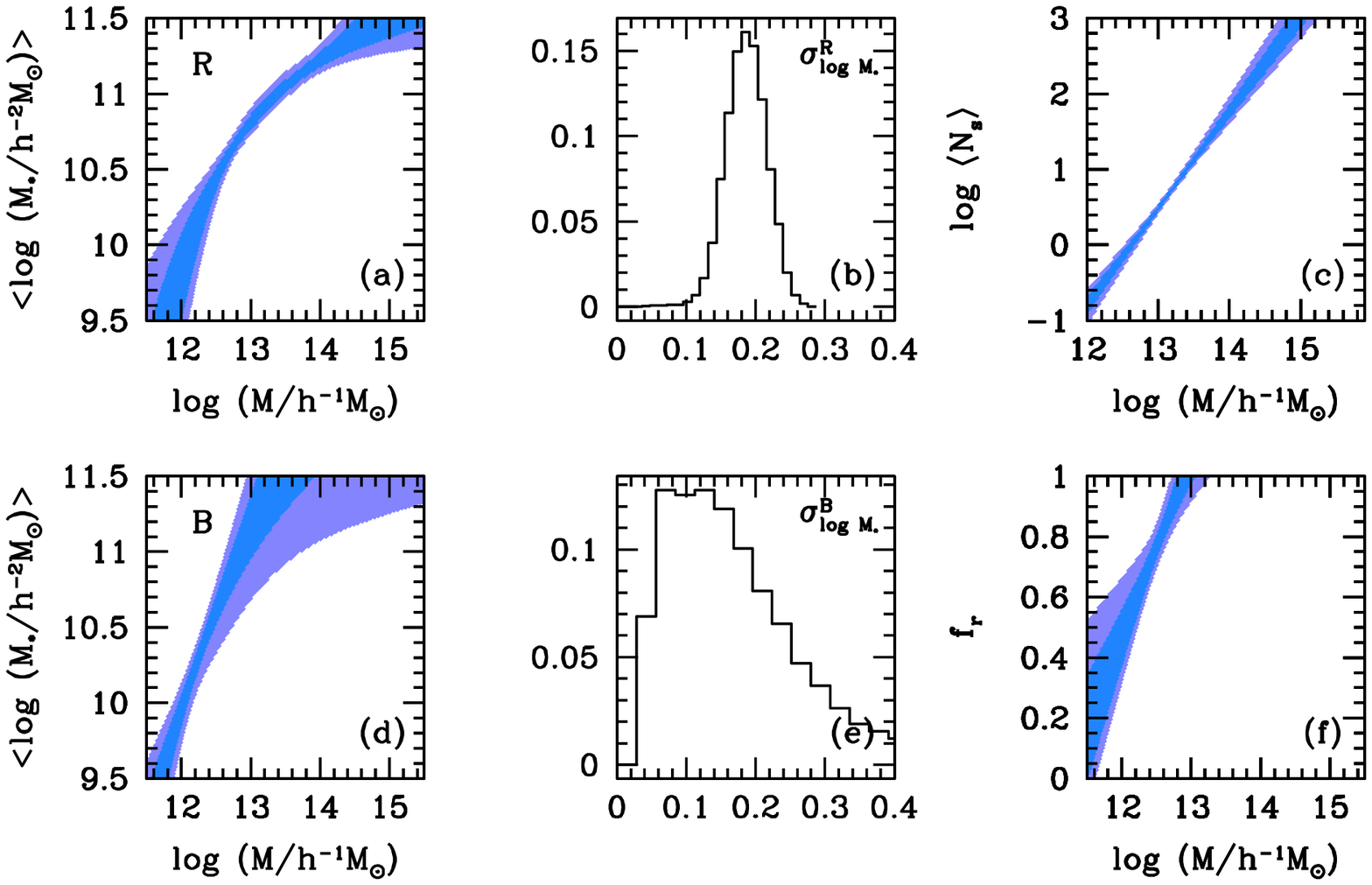,width=0.9\hdsize}}
\caption{Constraints on the model ingredients obtained from the MCMC
  The 68 and 95 percent confidence intervals are indicated as blue and
  purple regions, respectively. Panel (a) shows the average stellar
  mass of red centrals as a function of halo mass, while panel (b)
  shows the posterior distribution of the scatter $\sigmscen$ in this
  relation.  Panels (d) and (e) show the same but for blue
  centrals. Panel (c) shows the constraints on the average number of
  satellites as a function of halo mass, and finally, panel (f) shows
  the constraints on the fraction of red centrals as a function of
  halo mass.}
\label{fig:fig3_MSR}
\end{figure*}
\begin{figure*}
\centerline{\psfig{figure=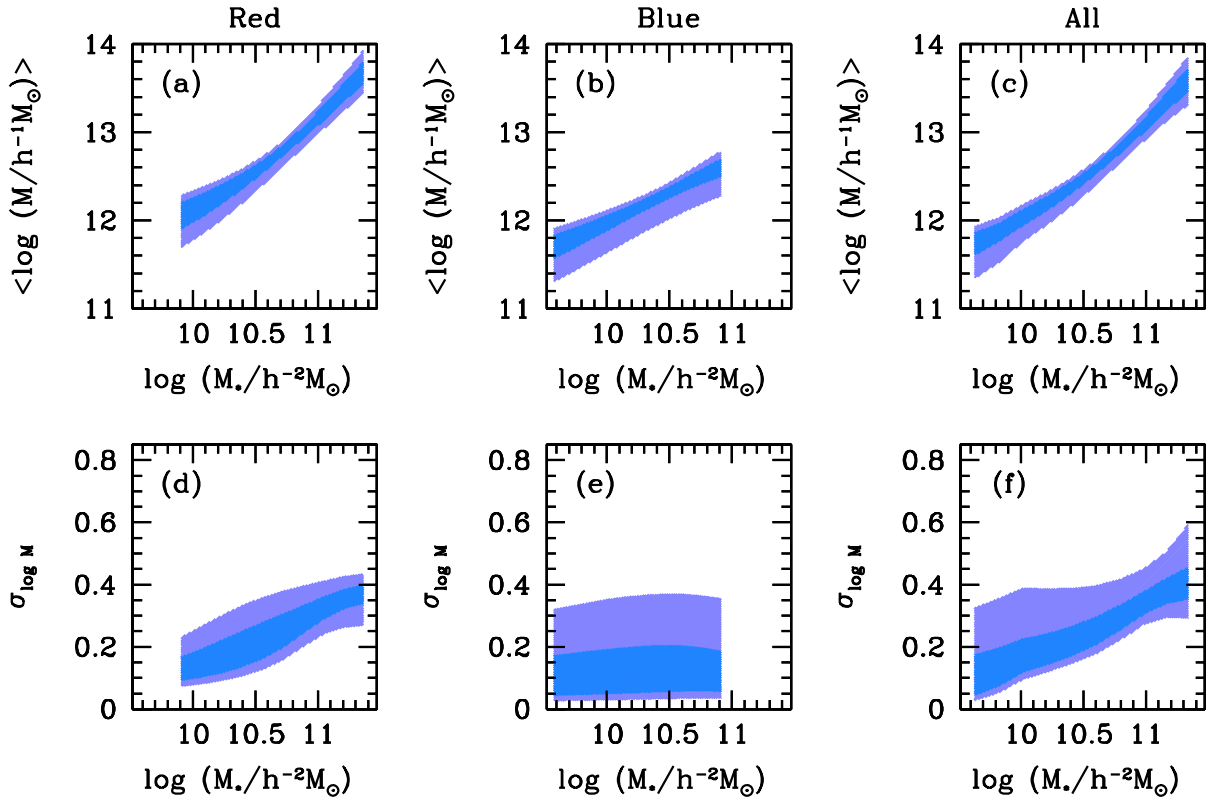,width=0.9\hdsize}}
\caption{Same as Fig.~\ref{fig:fig4_MLR}, but for the MSR, rather than
  the MLR.}
\label{fig:fig4_MSR}
\end{figure*}
\begin{figure*}
\centerline{\psfig{figure=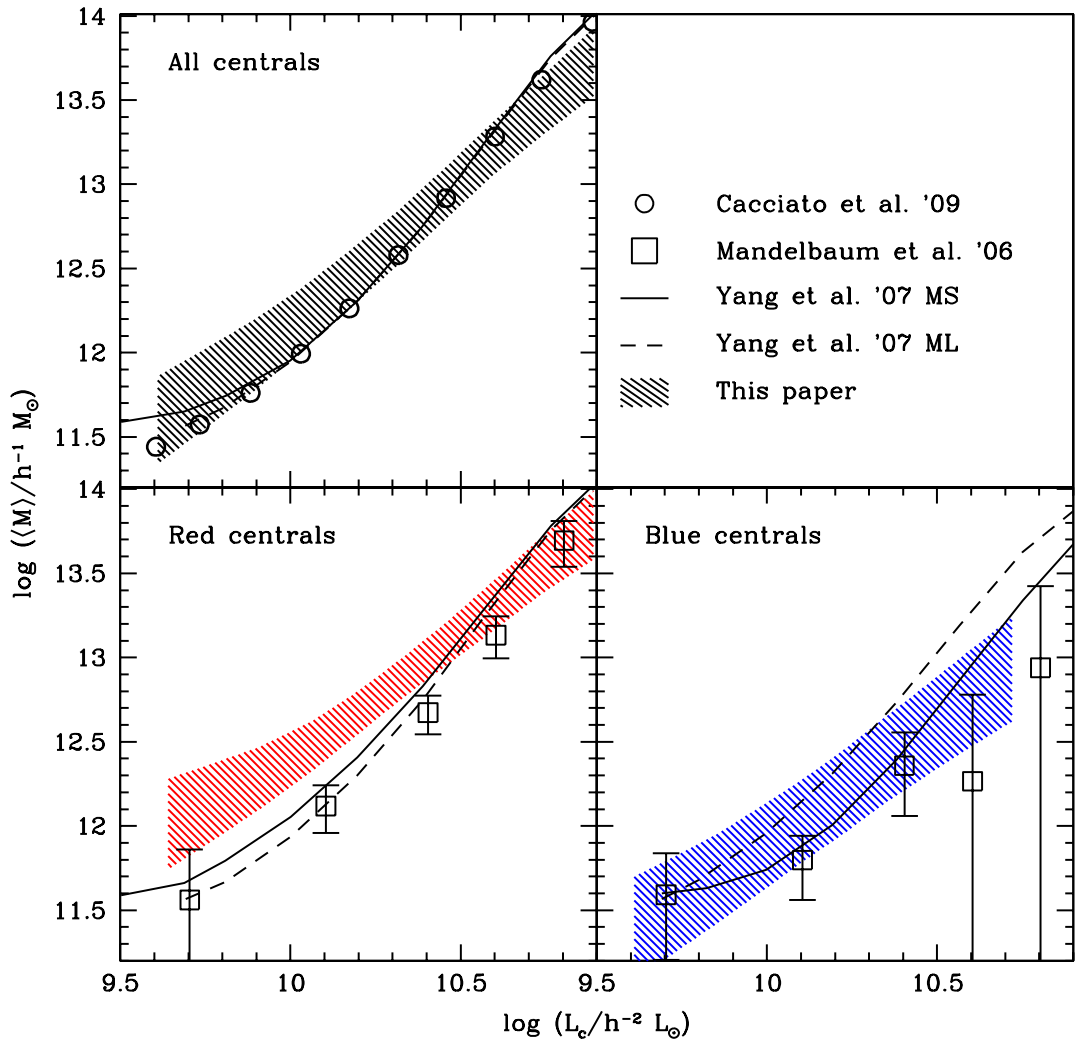,width=0.8\hdsize}}
\caption{Comparison of our MLR constraints with other constraints from
  the literature. The shaded regions show the 95 percent confidence
  intervals that we obtained from the analysis of satellite
  kinematics. The circles shows the MLR obtained by
  \citet{Cacciato2009} by using galaxy abundance and clustering
  measurements (the corresponding 95 percent confident intervals are
  smaller than the circles), the squares with errorbars (95 percent
  confidence intervals) show the MLR obtained by
  \citet{Mandelbaum2006} using weak lensing.  The solid and dashed
  lines show the MLR obtained from the group catalogue of
  \citet{Yang2007}.}
\label{fig:fig6_MLR}
\end{figure*}
\begin{figure*}
\centerline{\psfig{figure=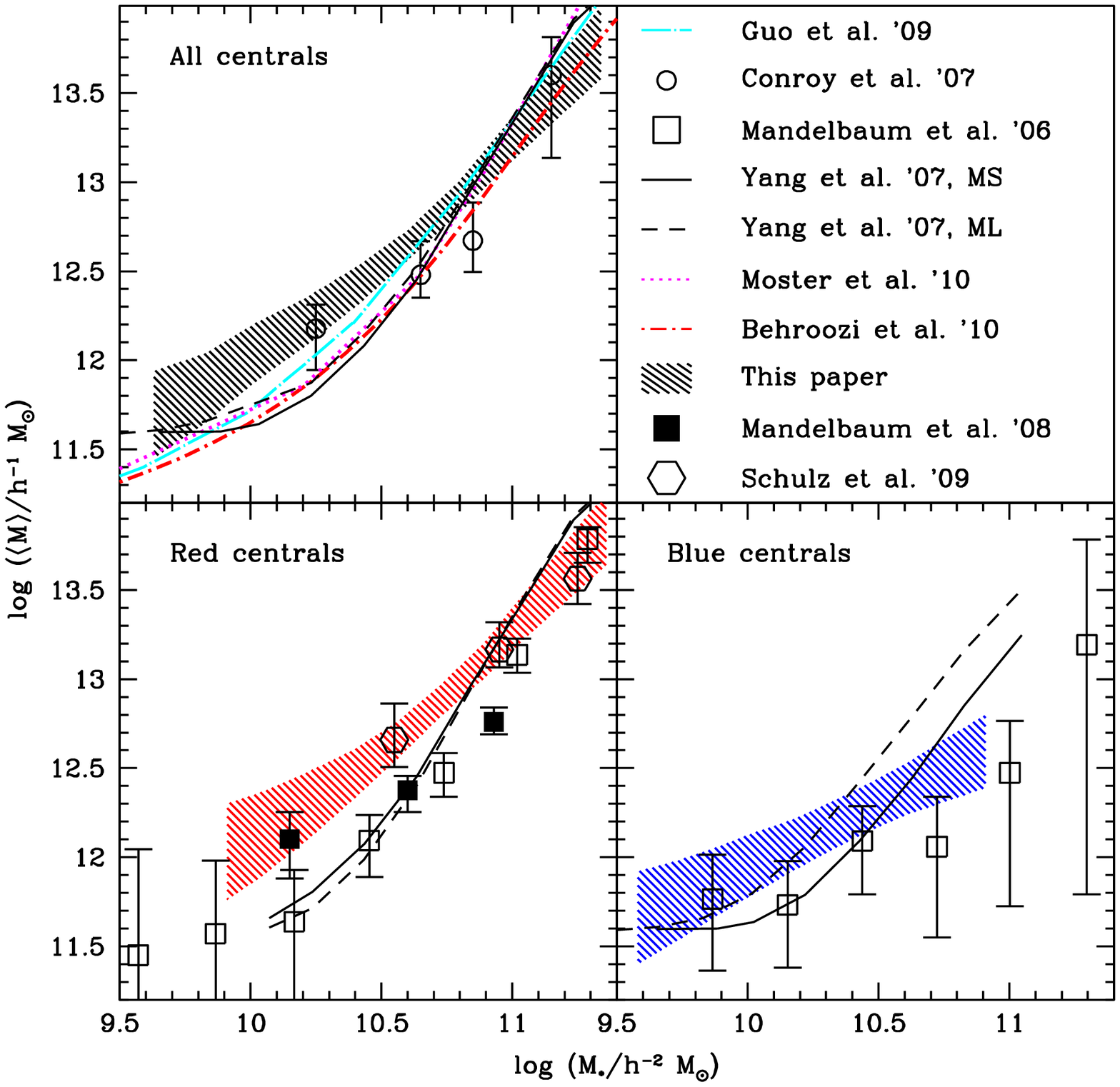,width=0.8\hdsize}}
\caption{Comparison of our MSR constraints with other constraints from
  the literature literature. The shaded regions show the 95 percent
  confidence intervals that we obtained from the analysis of satellite
  kinematics. The circles with errorbars (68 percent confidence
  intervals) shows the MSR obtained by \citet{Conroy2007} using
  satellite kinematics. The open squares, the solid squares and the
  hexagons with errorbars (95 percent confidence intervals) show the
  MSR obtained by \citet{Mandelbaum2006}, \citet{Mandelbaum2008} and
  \citet{Schulz2009} respectively, using weak lensing. The solid and
  dashed lines show the MSR obtained from the group catalogue of
  \citet{Yang2007}. The (magenta) dotted line, the (red) dot-dashed
  line and the (cyan) dot-long-dashed line show the results from
  abundance matching studies from \citet{Moster2010},
  \citet{Behroozi2010} and \citet{Guo2009} respectively. 
  }
\label{fig:fig6_MSR}
\end{figure*}

The analysis of the MSR of central galaxies is carried out by
analysing Samples SA, SR and SB (see Table~1 for the selection
criteria and the numbers of centrals and satellites in each of these
samples).  The host-weighted velocity dispersion, the
satellite-weighted velocity dispersion and the average number of
satellites as a function of the stellar mass of the central galaxies
obtained from Samples SR are shown as open squares in panels (a), (b)
and (c) of Fig.~\ref{fig:fig1_MSR}, respectively. In the same figure,
panels (d), (e) and (f) show the corresponding observables obtained
from Sample SB. At fixed stellar mass, the velocity dispersion of
satellite galaxies around red centrals is systematically larger than
that around blue centrals. The same is also true for the average
number of satellites. The fraction of red centrals as a function of
stellar mass obtained from Sample SA are shown in panel (g). The sharp
drop of $f_{\rm r}$ to zero at the low stellar mass end is due to
our use of a volume limited sample complete in luminosity, which
causes the sample selection function for red centrals to go to zero at
the low mass end (see Appendix~\ref{app:a}). The data in these 7
panels is used to constrain the 14 parameters of our model.  The blue
and purple shaded regions indicate the 68 and 95 percent confidence
levels obtained from our MCMC. As in the case of the MLR, the model
accurately fits the data, also in the case of sample SA, which is not
used to constrain the model (see Fig.~\ref{fig:fig2_MSR}). The 16, 50
and 84 percentile values of our model parameters obtained from our
analysis are listed in Table~\ref{tab:table3}.

Fig.~\ref{fig:fig3_MSR} is the same as Fig.~\ref{fig:fig3_MLR}, except
that it shows the constraints on the MSR, rather than the MLR.  At the
massive end, the average stellar mass of red centrals scales with halo
mass as $M_* \propto M^{0.29 ^{+0.15} _{-0.14} }$ while for blue
centrals we find that $M_* \propto M^{0.98 ^{+0.33} _{-0.50}}$. For
the scatter in $P(M_*|M)$ we obtain that $\sigmscen= 0.19 ^{+0.03}
_{-0.03}$ for red centrals and $\sigmscen= 0.15 ^{+0.12} _{-0.07}$ for
blue centrals. The posterior distribution of $\sigmscen$ for blue
centrals has a long tail extending to large values of scatter.  At the
low end, the confidence levels for $\sigmscen$ are largely dominated
by our prior ($\sigmscen > 0.04$), which has been adopted for
computational convenience. Hence, we cannot rule out that the data is
consistent with zero scatter (i.e. $\sigmscen = 0.0$) for blue
centrals stacked by stellar mass.

%it is unclear whether the data is
%consistent with zero scatter (i.e.  $\sigmscen = 0.0$) or not. 

In Fig.~\ref{fig:fig4_MSR} we present the average halo mass and the
scatter in halo masses obtained from our analysis for red centrals
(left-hand panels), blue centrals (middle panels) and all centrals
(right-hand panels) as a function of stellar mass. Note that the
scatter in halo masses increases with stellar mass for the red
centrals, while it is roughly independent of stellar mass for blue
centrals.  In the right-hand panel of Fig.~\ref{fig:fig5_MLRMSR} we
compare the MSR of red centrals (triangles) to that of blue centrals
(circles) and to that of all centrals (squares). At $M_* \lta
10^{10.5}\Msunhh$, the average halo mass of red centrals is virtually
identical to that of blue centrals, certainly within the errorbars. At
larger stellar masses, red centrals reside on average in more massive
haloes than blue centrals. In contrast, at fixed luminosity (see
left-hand panel), the average halo mass of red centrals is {\it
always} systematically larger that that of blue centrals, by more than
a factor of three. Hence, we conclude that stellar mass is a better
indicator of halo mass than luminosity, especially at the low mass
end. However, we also stress that there still is a significant amount 
of scatter in the relation between the stellar mass of a central
galaxy and the mass of the halo in which it resides.

%Finally, we would like to mention that our fiducial model assumes that
%the number of satellites in a given halo, $\avnsat_{M}$, depends only
%on the mass of the halo. We have also carried out the analysis of the
%MLR and MSR using a model in which we relaxed this assumption. In this
%model, we allowed $\avnsat_{M}$ to also depend on the colour of the
%central galaxy. We find that the $\avnsat_{M}$ around red and blue
%centrals is consistent with being the same. We also find that the
%resulting constraints on the MLR and MSR are slightly weaker,
%especially for the blue galaxies but are not significantly different
%from those inferred by using the fiducial model.

Finally, recall that we have assumed that the number of satellites in
a given halo, $\avnsat_{M}$, is independent of the color of its
central. In order to test the possible implications of this
assumption, we have repeated the above analysis allowing for
independent $\avnsat_{M}$ for haloes with red and blue centrals (each
parameterized with Eq.~[\ref{nsatm}]). This adds an additional two
free parameters to the model, bringing the total to 16.  We find that
the resulting constraints are perfectly consistent with $\avnsat_{M}$
being identical for red and blue centrals, and that all other
constraints are similar to what we presented above. Hence, we conclude
that our assumption that $\avnsat_{M}$ is independent of the color of
its central is supported by the data and does not bias our results.

\subsection{Comparison with other studies}
\label{sec:comp}

It is interesting and important to compare our constraints on the MLR
and MSR of central galaxies to those obtained using different,
independent data sets and methods, including galaxy group catalogues,
galaxy-galaxy lensing, galaxy clustering, halo abundance matching, and
other studies of satellite kinematics. In order to enable a fair and
meaningful comparison, whenever required we have adapted the results
in the literature to match the definitions of halo mass and stellar
mass used in this paper. In particular, we follow \citet{Hu2003} to
convert between different definitions of halo mass and use the results
of \citet{Bell2003} and \citet{Borch2006} to convert stellar masses to
the \citet{Kroupa2001} initial mass function adopted here.

\citet{Yang2007} studied galaxy groups in the SDSS, which they
assigned halo masses based upon either the total stellar mass or the
total luminosity content of each group. We use their group catalogue to
investigate the MLR and MSR of central galaxies, with and without the
split in red and blue by colour. The solid and dashed lines in
Figs.~\ref{fig:fig6_MLR} and~\ref{fig:fig6_MSR} correspond to the MLR
and MSR of central galaxies in their group catalog, where the halo
masses have been assigned using the total stellar mass and total
luminosity content of the groups, respectively.

\citet{Mandelbaum2006} measured the galaxy-galaxy lensing signal in
the SDSS for galaxies stacked by luminosity and by stellar mass. The
galaxies were split into early and late types based upon their
morphology. Here we make the crude assumption that this is equivalent
to our split in red and blue; although clearly an oversimplification,
we do not believe that this invalidates a comparison with our results.
Modelling their data, they obtained the MLR and MSR indicated in
Figs.~\ref{fig:fig6_MLR} and \ref{fig:fig6_MSR} by open squares with
errorbars. The same exercise was repeated in \citet{Mandelbaum2008},
but by stacking isolated galaxies. The results of this analysis are
shown as filled squares with errorbars (95 percent) in
Fig.~\ref{fig:fig6_MSR}. In the same figure, we also show the
galaxy-galaxy lensing results of \citet{Schulz2009} as hexagons with
errorbars (95 percent).

The abundance and clustering of galaxies holds important information
regarding the halo occupation statistics of these galaxies. Using the
observed luminosity function and clustering properties of galaxies in
the SDSS, \citet{Cacciato2009} constrained the conditional luminosity
function, $\Phi(L|M)$, which describes the average number of galaxies
of luminosity $L$ that reside in a halo of mass $M$
\citep[][]{Yang2003}.  We use their best fit parameters to calculate
the MLR of central galaxies and show it using open
circles\footnote{\citet{Cacciato2009} also used results from the group
catalogue of \citet{Yang2007} to constrain their model, which explains
why (i) their errorbars are extremely small and (ii) their results are
in perfect agreement with those of \citet{Yang2007}} in the top
left-hand panel of Fig.~\ref{fig:fig6_MLR}. Note that
\citet{Cacciato2009} have also shown that their halo occupation model
is able to reproduce the galaxy-galaxy lensing signal obtained by
stacking central galaxies based upon their luminosities.

The relation between galaxy stellar mass and halo mass can also be
obtained by matching the abundance of galaxies to the abundance of
haloes and subhaloes that host these galaxies
\citep[e.g.][]{Conroy2006, Conroy2009, Guo2009, Moster2010,
  Behroozi2010}. The results from this (sub)halo abundance matching
technique are often quoted as the mean of the distributions $P(\log
M_*|M)$. We asked the respective authors to provide us $\log
\avg{M}(M_*)$ to enable a fair comparison. In the top left panel of
Fig.\ref{fig:fig6_MSR} we show the results of \citet[][dotted
line]{Moster2010}, \citet[] [dot-long-dashed line]{Guo2009} and
\citet[][dot-dashed line]{Behroozi2010}.

Finally, using data from the SDSS and the DEEP2 survey,
\citet{Conroy2007} used the kinematics of satellite galaxies to
determine the evolution of the stellar mass-to light ratio of central
galaxies from $z\sim1$ to $z\sim0$. They measured and modelled the
radial dependence of the velocity dispersion (in contrast to the
aperture averaged velocity dispersions used in this paper) to infer
the average halo mass as a function of the stellar mass of the central
galaxy. The halo mass$-$stellar mass relation for all central galaxies
thus obtained from their analysis at $z\sim0$ is shown in
Fig.~\ref{fig:fig6_MSR} using circles with errorbars. 

In all panels of Figs.~\ref{fig:fig6_MLR} and \ref{fig:fig6_MSR}, the
shaded areas show the 95 percent confidence intervals obtained in this
paper using the kinematics of satellite galaxies. Overall, the results
obtained using all these very different methods are in remarkably good
agreement with each other \citep[see also][]{Behroozi2010,
Dutton2010}, and with the results obtained here. However, there also
are a few discrepancies, which we discuss below.

With regard to the MLR, the galaxy group catalogues of
\citet{Yang2007} and the galaxy-galaxy lensing analysis of
\citet{Mandelbaum2006} yield halo masses around red centrals that are
in good agreement with each other. But at $L \sim 10^{10}
h^{-2}\Lsun$, the halo masses inferred from these two methods are
roughly a factor two lower compared to the results obtained here. The
three methods, however, agree fairly well at the bright end. Somewhat
surprisingly, the MLR of all centrals (upper left-hand panel of
Fig.~\ref{fig:fig6_MLR}) does not reveal any discrepancy between the
masses inferred from satellite kinematics, versus those inferred from
either clustering \citep[results of][]{Cacciato2009} or galaxy group
catalogues \citep[results of][]{Yang2007}.  In case of the MSR of red
centrals, at the low stellar mass end the group catalogue results
agree with the weak lensing results and are again a factor two to
three lower than the results obtained here. On the other hand, at the
bright end, our results agree fairly well with the weak lensing
results while the group catalogue results are roughly larger by a
factor of 1.6. It is also worth noting that the weak lensing results
are not fully consistent among each other at the intermediate and low
mass ends. In particular, our results are in excellent agreement with
the weak lensing analysis of \citet{Schulz2009} and the low mass point
of \citet{Mandelbaum2008}. For the MSR of all centrals (upper
left-hand panel of Fig.~\ref{fig:fig6_MSR}), our analysis of satellite
kinematics once again yields halo masses around low-mass centrals that
are $\sim 0.3$ dex larger than those inferred using either subhalo
abundance matching or galaxy group catalogues.  It is noteworthy,
though, that the results obtained by \citet{Conroy2007}, which are
also based on satellite kinematics, are actually in good agreement
with our results.  Finally, we note that for blue centrals there is no
clear indication of any systematic discrepancy, except perhaps at the
massive end.  However, since the massive (bright) end of the galaxy
mass (luminosity) function is dominated by red centrals, the
corresponding number statistics are poor resulting in large
errorbars. Consequently, we do not consider this discrepancy
significant.

To summarize, Figs.~\ref{fig:fig6_MLR} and~\ref{fig:fig6_MSR} indicate
that tremendous progress has been made in recent years in constraining
the galaxy-dark matter connection, with different techniques yielding
MLRs and MSRs that are in fairly good agreement with each other,
typically within a factor of two. While it is difficult to make
any robust statement about possible systematics, we acknowledge that
there is a hint that satellite kinematics yields halo masses around
low mass centrals that are systematically larger than most other
methods, especially around red centrals. Although we certainly can't rule
out any systematics in the other methods, we briefly discuss a
potential problem with satellite kinematics.

Recently, \citet{Skibba2010} analyzed the SDSS galaxy group catalogue
of \citet{Yang2007} and showed that in a significant fraction of
groups (ranging from $\sim 25$ percent at the low mass end to $\sim
40$ percent at the massive end) the brightest group member is a
satellite galaxy rather than a central.  As discussed at length in
their paper, this could cause satellite kinematics to overestimate
halo masses by as much as a factor of $\sim 1.6$. However, we consider
it unlikely that this explains the systematic offset between satellite
kinematics and other methods because of the following two reasons:
First of all, the effect is expected to be largest at the massive end,
and to be negligible at the low mass end, opposite to the trends seen
in Figs.~\ref{fig:fig6_MLR} and~\ref{fig:fig6_MSR}.  Secondly, the
factor $\sim 1.6$ overestimate only occurs if the probability ${\cal
P}_{\rm BNC}(M)$ that the brightest (most massive) galaxy in a halo of
mass $M$ is not the central galaxy is independent of the luminosity
(stellar mass) of the central galaxy. However, if haloes of mass $M$
in which the central galaxy is fainter (less massive) than the average
for haloes of that mass are more prone to having a brighter (more
massive) satellite, which does not seem unreasonable, then the effect
can be much weaker \citep[see][for details]{Skibba2010}.

\begin{figure}
\centerline{\psfig{figure=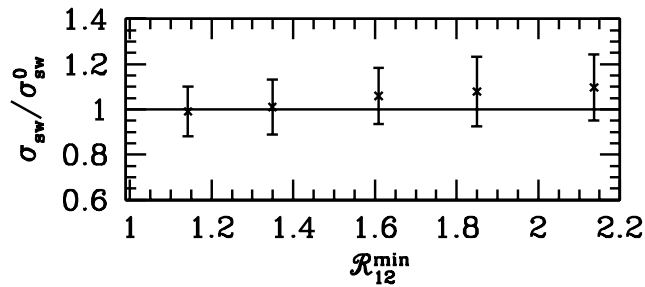,width=\hssize}}
\caption{The ratio of the satellite-weighted velocity dispersion,
$\sigma_{\rm sw}$, measured around hosts that have ${\cal R}_{12}$
(the ratio of their own stellar mass to that of their most massive
satellite) greater than or equal to ${\cal R}^{\rm min}_{12}$, to the
ratio of the satellite-weighted velocity dispersion measured around
all hosts ($\sigma^0_{\rm sw}$).}
\label{fig:fig7}
\end{figure}
To investigate the possibility that our results for red centrals at
the low stellar mass end are affected by haloes in which the central
galaxy is not the most massive, we perform the following test. We take
all hosts from Sample SR that have stellar masses in a bin of width
$0.16$ dex centered around $\log(M_*/\Msunhh)=10.39$.  For each host,
we determine the ratio, ${\cal R}_{12}$, of its stellar mass to that
of its most massive satellite. Under the assumption that a larger
${\cal R}_{12}$ implies a larger probability that this host is a true
central, we proceed as follows. We sort the hosts in decreasing order
of ${\cal R}_{12}$, and create 5 samples by discarding the final $10$,
$20$, $30$, $40$ and $50$ percent of the hosts (plus their
satellites), respectively. For each of these samples we compute the
satellite-weighted velocity dispersion of the remaining satellites,
which we plot in Fig.~\ref{fig:fig7} as a function of the minimum
value of ${\cal R}_{12}$. Clearly, within the errorbars there is no
indication for a systematic trend between $\sigma_{\rm sw}$ and ${\cal
R}^{\rm min}_{12}$. Although this does not rule out that our satellite
kinematics are affected by the possibility that a certain fraction of
our host galaxies in reality are satellites rather than centrals, it
certainly makes it less likely.  

\section{Summary}
\label{sec:summary}

We have used the kinematics of satellite galaxies in order to probe
the halo mass-luminosity relation (MLR) and the halo mass-stellar mass
relation (MSR) of central galaxies. For this purpose, an iterative
selection criterion was first applied to select central and satellite
galaxies from a volume-limited sub-sample of the SDSS.  The resulting
sample consist of $\sim 6200$ satellites around $\sim 3900$ centrals,
making it the largest volume-limited sample of central-satellite
pairs used to date for studies of satellite kinematics. Since the
number of satellite galaxies around any individual central galaxy is
small, a stacking procedure was used to combine the velocity
information of the satellite galaxies of centrals with similar
luminosities or stellar masses. A detailed modelling procedure,
outlined in Paper I, was then used to infer both the average scaling
relation between halo mass and the central galaxy property and its
scatter.

As expected, and in qualitative agreement with many other studies, we
find that more luminous (more massive) centrals reside in more massive
haloes. In addition, we find that the MLR of central galaxies is
different for central galaxies of different colour: red centrals, on
average, occupy more massive haloes than blue centrals of the same
luminosity. Consequently, the scatter in the MLR of central galaxies
is at least partly correlated with the colour of the central galaxy. 

When stacking central galaxies according to their stellar masses, we
find the difference in the mean MSRs of red and blue centrals to be
less pronounced than in the case of the MLR.  In particular, for $M_*
\lta 10^{10.5}\Msunhh$, the average halo masses of red and blue
centrals are not significantly different. We thus conclude that the
stellar mass of a central galaxy is a more reliable indicator of its
halo mass than its ($r$-band) luminosity. However, even the MSR has a
significant amount of scatter of the order of $\sim 0.2$ dex in stellar
mass at fixed halo mass. This translates into a scatter in halo mass at
fixed stellar mass that increases from $\sim 0.1$ dex at $M_* \simeq 4
\times 10^9 \Msunhh$ to $\sim 0.4$ dex at $M_* \simeq 2 \times 10^{11}
\Msunhh$.

We compared our constraints on the MLR and MSR of central galaxies
with a number of other, independent constraints coming from the SDSS
galaxy group catalogue of \citet{Yang2007}, the galaxy-galaxy lensing
analyses of \citet{Mandelbaum2006}, \citet{Mandelbaum2008} and
\citet{Schulz2009}, the galaxy clustering analysis of
\citet{Cacciato2009}, subhalo abundance matching studies of
\citet{Moster2010}, \citet{Guo2009} and \citet{Behroozi2010}, and the
analysis of satellite kinematics by \citet{Conroy2007}. Overall, there
is good agreement among all these different studies, with a typical
study-to-study scatter of less than a factor two, which is comparable
to the typical 2$\sigma$ errors quoted by most of these studies.
There is some indication, though, that satellite kinematics yield halo
masses around low-mass centrals ($M_* \lta 3 \times 10^{10}\Msunhh$)
that are systematically higher by a factor 2-3 that most other
methods, although we emphasize that our results are in perfect
agreement with the galaxy-galaxy lensing analysis of
\citet{Schulz2009}. As discussed in the text, we do not believe that
our results are significantly affected by the fact that not all
central galaxies are the brightest (or most massive) galaxies in their
dark matter haloes, as shown by \citet{Skibba2010}. In fact, detailed
tests with mock galaxy redshift surveys, presented in Paper~II, have
revealed no systematic effects for our method of analysis.

We conclude that the overall level of agreement regarding the MLR and
MSR among all different techniques indicates that we are converging on
an accurate and reliable description of the galaxy-dark matter
connection \citep[see also][]{Bosch2007,Behroozi2010,Dutton2010}. In
addition to an overall agreement regarding the means of the MSR and
MLR, to well within a factor of two, there is also good agreement
regarding the amount of scatter; as demonstrated in Paper~II, the
scatter of $\sim 0.2$ dex in luminosity or stellar mass at fixed halo
mass is in excellent agreement with independent constraints obtained
by \citet{Yang2008}, \citet{Cacciato2009} and \citet{Cooray2006}, and
with predictions from the semi-analytical model for galaxy formation
of \citet{Croton2006}. This overall level of agreement is an admirable
achievement, which will prove invaluable not only for furthering our
understanding of galaxy formation, but also for using galaxies to
probe the cosmic density field and to constrain cosmological
parameters.

\section{Acknowledgments} 

It is a pleasure to thank Eric Bell, Kris Blindert and Anupreeta More
for useful discussions, and Qi Guo, Peter Behroozi and Benjamin
Moster for providing results from their abundance matching. For a
significant duration of this project, SM and MC were members of the
IMPRS for Astronomy \& Cosmic Physics at the University of Heidelberg.

\appendix

\section{Sample Selection Function}

\label{app:a}
\begin{figure*}
\centerline{\psfig{figure=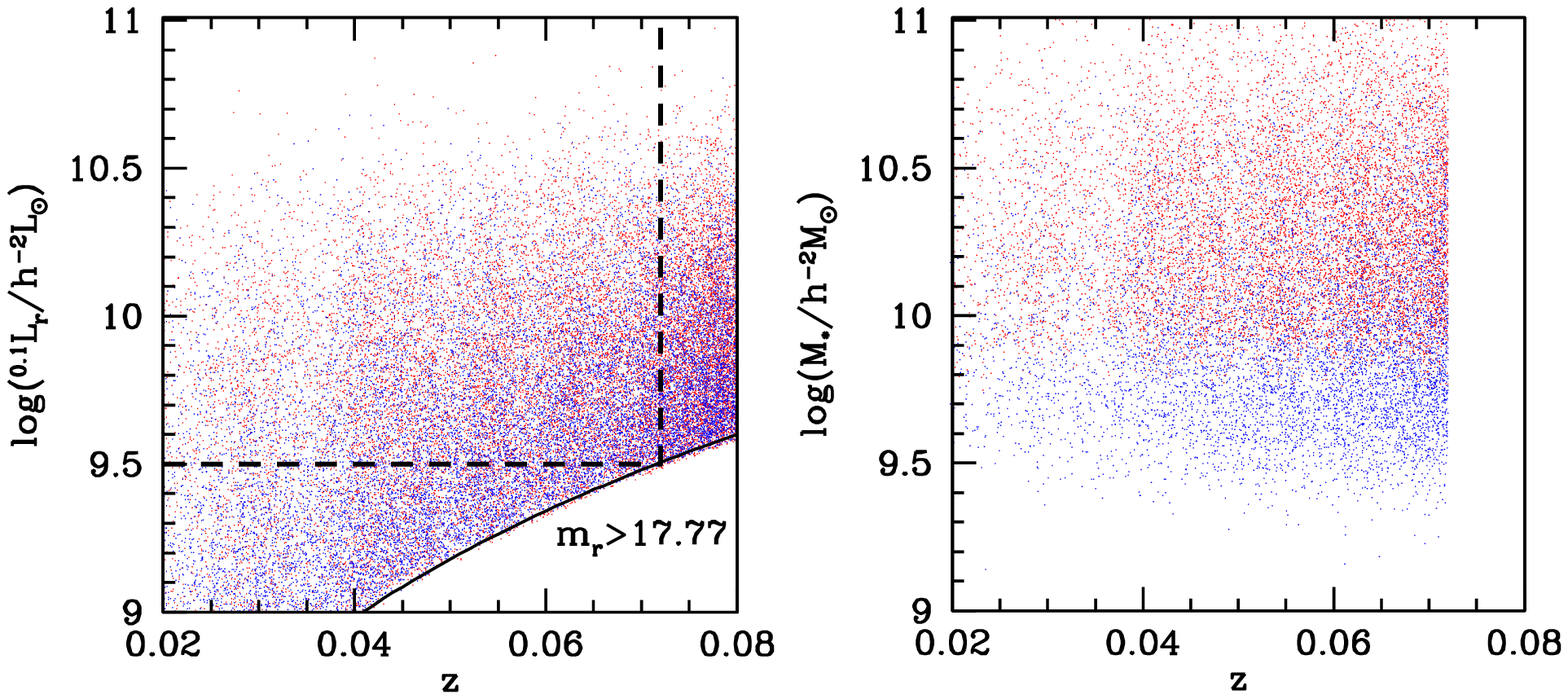,width=0.9\hdsize}}
\caption{The left panel shows the distribution of SDSS galaxies in
  the luminosity-redshift plane. Red and blue galaxies are indicated
  by red and blue dots, respectively. The volume-limited sample of
  galaxies used for our analysis of satellite kinematics is enclosed
  by the dashed lines. The right hand panel shows the
  distribution of galaxies in this volume-limited sample in the stellar
  mass-redshift plane.}
\label{fig:fig1_App}
\end{figure*}

Here we illustrate how our use of a volume limited sample of galaxies,
complete in luminosity, leads to a sample that is incomplete in
stellar mass. Our aim is to characterize the sample selection
function, $S(M_*,C)$, which describes the fraction of galaxies of
stellar mass $M_*$ and colour $C$ that make it into our sample. This
sample selection function is required for modelling the satellite
kinematics as function of the stellar mass of their host galaxy 
(see Section~\ref{sec:hos})

The left-hand panel of Fig.~\ref{fig:fig1_App} shows the distribution
of galaxies in the $^{0.1}r$-band luminosity-redshift plane. Red and
blue galaxies are indicated by red and blue dots, respectively. The
apparent magnitude limit, $m_r^{\rm lim}=17.77$, of the spectroscopic
sample results in an absolute magnitude limit given by
\begin{eqnarray}
\lefteqn{^{0.1}M_r^{\rm lim}-5 \log h = 17.77 - {\rm DM}(z) -
  k_{0.1}(z)} \nonumber \\
& & + 1.62\,(z-0.1)\,.
\end{eqnarray}
Here $k_{0.1}(z)$ is the k-correction to redshift $z=0.1$, $1.62$ is
the evolution correction factor from \citet{Blanton2003a}, and DM($z$)
is given by
\begin{equation}
{\rm DM}(z) = 5 \log D_L (z) + 25.0\,,
\end{equation}
where $D_L(z)$ is the luminosity distance of the galaxy in
$\mpch$. The redshift dependence of the k-corrections is fairly well
reproduced by \citep{Bosch2008}
\begin{equation}
\label{kpone}
k_{0.1}(z)=2.5\log\left(\frac{z+0.9}{1.1}\right)\,.
\end{equation}
The above equations yield the solid black line in the left-hand panel
of Fig.~\ref{fig:fig1_App}. Note that a very small fraction of
galaxies fall below this predicted limit. This is because the
k-correction also depends on the colour of the central galaxy which we
have not accounted for. However this effect can safely be ignored for
the purpose of this paper.

Galaxies with $z \leq 0.072$ and $^{0.1}L_r \geq 10^{9.5}$ (i.e.,
$^{0.1}M_r^{\rm lim}-5 \log h \leq -18.99$) make up the volume limited
sample used in this paper (indicated by dashed lines in the left-hand
panel of Fig.~\ref{fig:fig1_App}). The right-hand panel of
Fig.~\ref{fig:fig1_App} shows the distribution of galaxies in this
volume limited sample in the stellar mass-redshift plane. Note that
the sharp cut in luminosity translates into a colour-dependent cut in
stellar mass; bluer galaxies have a lower cut-off in stellar mass,
causing the low-mass end of the sample to be completely dominated by
blue galaxies.

We remind the reader that in our analysis, galaxies were assigned
stellar masses using the $^{0.0}(g-r)$ colour and the $^{0.0}r$-band
magnitude using Eq.~(\ref{eq:bell2003}). The $^{0.1}r$ band magnitude
limit we have used for our volume limited sample translates into a
$^{0.0}r$-band limit given by
\begin{equation}
\label{zerotopone}
^{0.0}M_r^{\rm lim} = ^{0.1}M_r^{\rm lim} + [k_{0.1}(z) - k_{0.0}(z)] - 0.162\,,
\end{equation}
where the term in square brackets is the difference in k-corrections
between redshift $0.0$ and $0.1$, and the last term is the difference
in the evolution corrections between these redshifts. The
k-corrections to redshift $0.0$ can be reasonably well described by
\citep{Bosch2008}
\begin{equation}
\label{kzero}
k_{0.0}(z)=2.5\log(1+z)+1.5\,z\,[^{0.0}(g-r)-0.66]\,
\end{equation}
Combining Eqs.~(\ref{zerotopone}), (\ref{kzero}), (\ref{kpone}) with
Eq.~(\ref{eq:bell2003}) gives that, at fixed $^{0.0}(g-r)$ colour, the
stellar mass limit in the volume limited sample varies with
$^{0.0}(g-r)$ and redshift as
\begin{eqnarray}
\lefteqn{\log(M_*^{\rm lim}) = 8.9812 -
\log\left[\frac{z+0.9}{1.1\,(1+z)}\right]} \nonumber \\ 
& & - 0.396\,z + (1.097+0.6\,z\,)^{0.0}(g-r)
\label{eq:climit}
\end{eqnarray}

\begin{figure}
\centerline{\psfig{figure=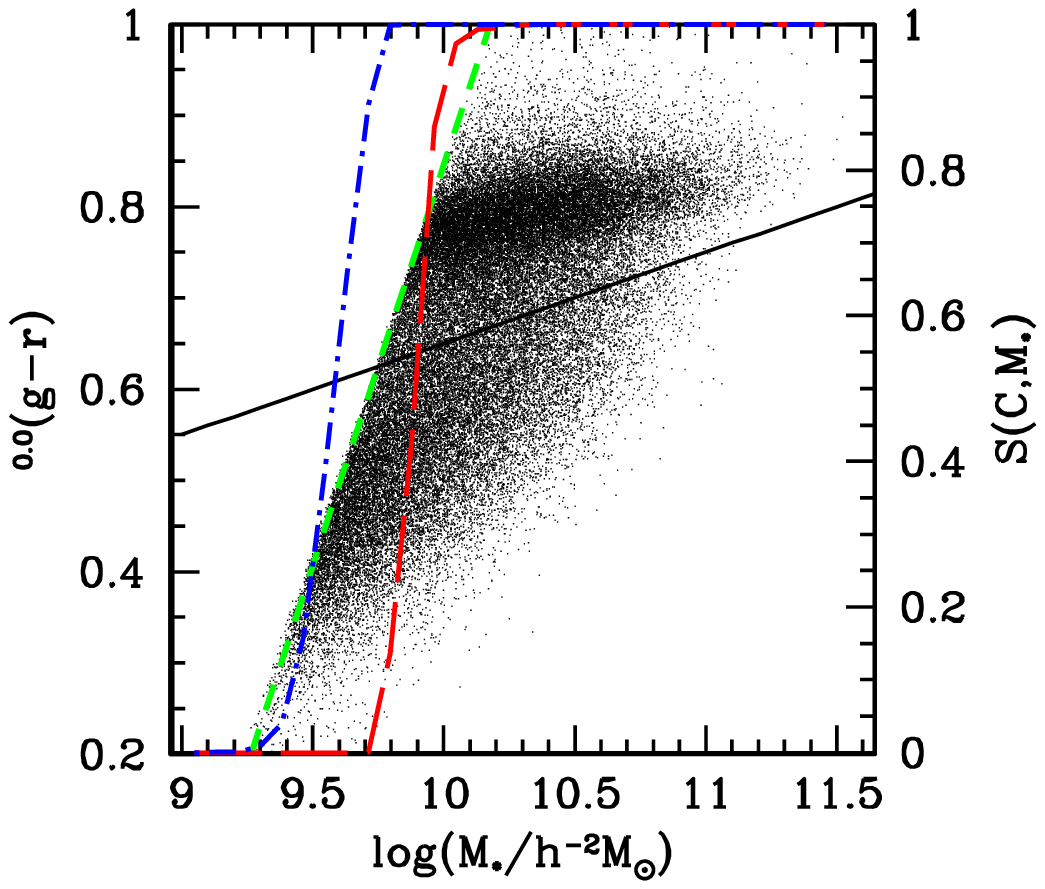,width=\hssize}}
\caption{The colour-stellar mass diagram of galaxies in our
  volume-limited sample. The black solid line shows the boundary that
  we use to split the galaxy population in `red' and `blue'
  galaxies. The green dashed line shows the analytical prediction
  (Eq.~[\ref{eq:climit}]) for the selection effect that results due to
  our use of a volume-limited sample complete in luminosity.  The
  resulting selection functions, $S(C,M_*)$, for red and blue galaxies
  are indicated by the (red) dashed and (blue) dot-dashed lines,
  respectively.}
\label{fig:fig2_App}
\end{figure}

In Fig.\ref{fig:fig2_App}, we plot the distribution of galaxies in the
colour-stellar mass diagram. The colour cut which was used to assign
colours to our sample of galaxies is shown by the black solid line
(see Eq.~[\ref{eq:bell2003}]). The green dashed line show the limit
expressed in Eq.~(\ref{eq:climit}) assuming $z=0.072$. At fixed stellar
mass, only a fraction of galaxies (those that lie below the green
line), are part of the volume limited sample. More than half of the
red galaxies drop out of the sample at stellar masses below
$10^{9.8}\Msunhh$.

To calculate the selection function $S(M_{*},C)$, we use the entire
flux limited catalogue of galaxies to populate the
$^{0.0}(g-r)-$stellar mass plane, this time assigning each galaxy a
weight equal to one over the maximum volume to which this galaxy could
be seen given the $r$-band flux limit of $17.77$. Using small bins in
stellar mass, we calculate $S(M_{*},C)$ in each bin by dividing the
sum of weights of galaxies of a particular colour that lie below
$\log(M_*^{\rm lim})$ with the sum of weights of all galaxies of that
colour in the bin under consideration. The resulting selection
functions for red and blue galaxies are plotted in
Fig.~\ref{fig:fig2_App} as red, long-dashed and blue dot-dashed lines,
respectively.

\end{document}

%% file: macros.tex
\newcommand{\etal}{{et al.~}}

\newcommand{\kms}{\>{\rm km}\,{\rm s}^{-1}}

\newcommand{\Mpc}{\>{\rm Mpc}}

\newcommand{\Msun}{\>{\rm M_{\odot}}}
\newcommand{\Lsun}{\>{\rm L_{\odot}}}

\newcommand{\mpch}{\>h^{-1}{\rm {Mpc}}}

\newcommand{\Msunh}{\>h^{-1}\rm M_\odot}

\newcommand{\beq}{\begin{equation}}
\newcommand{\eeq}{\end{equation}}

\newcommand{\rmd}{{\rm d}}

\newcommand{\Lsunhh}{\,h^{-2}\rm L_\odot}
\newcommand{\Msunhh}{\,h^{-2}\rm M_\odot}

\newcommand{\lcen}{L}

\newcommand{\pc}{Q}
\newcommand{\pten}{Q_{\rm 10}}
\newcommand{\lc}{\lcen}
\newcommand{\avg}[1]{\langle #1 \rangle}

\newcommand{\drm}{{\rm d}}
\newcommand{\pdv}{{P (\Delta V)}}
\newcommand{\dv}{{\Delta V}}
\newcommand{\siglcen}{\sigma_{\log L}}
\newcommand{\sigcenm}{\sigma_{\log M_*}}
\newcommand{\sigmscen}{\sigma_{\log M_*}}

\newcommand{\sigsw}{\sigma_{\rm sw}}
\newcommand{\sighw}{\sigma_{\rm hw}}
\newcommand{\sigsat}{\sigma_{\rm sat}}
\newcommand{\avgsigsatsq}{\avg{\sigsat^2}}
\newcommand{\avnsat}{\avg{N_{\rm s}}}
\newcommand{\avnsatm}{\avnsat(M)}

\newcommand{\plcm}{P(\lcen|M)}

\newcommand{\pmpc}{P(M|\pc)}

\newcommand{\Rh}{R_{\rm h}}
\newcommand{\Rs}{R_{\rm s}}
\newcommand{\dvh}{(\dv)_{\rm h}}
\newcommand{\dvs}{(\dv)_{\rm s}}
\newcommand{\Sset}{\hat{S}}
\def\gtsima{$\; \buildrel > \over \sim \;$}
\def\ltsima{$\; \buildrel < \over \sim \;$}
\def\prosima{$\; \buildrel \propto \over \sim \;$}
\def\gsim{\lower.7ex\hbox{\gtsima}}
\def\lsim{\lower.7ex\hbox{\ltsima}}
\def\simgt{\lower.7ex\hbox{\gtsima}}
\def\simlt{\lower.7ex\hbox{\ltsima}}
\def\simpr{\lower.7ex\hbox{\prosima}}
\def\la{\lsim}

\def\lta{\la}

\newcommand{\apj}{ApJ}

\newdimen\hssize
\newdimen\hdsize